\documentclass[lettersize,journal]{IEEEtran}
\usepackage{color,array,amsthm}
\usepackage{graphicx}

\usepackage{amsmath}
\usepackage{amssymb}

\usepackage{subcaption}

\usepackage[numbers,sort&compress]{natbib}

\usepackage{hyperref}
\hypersetup{colorlinks=true, linkcolor=blue, anchorcolor=blue, citecolor=blue}
\usepackage{cleveref}

\usepackage{float}
\usepackage{booktabs}
\usepackage{multirow}

\usepackage{algorithm}
\usepackage{algorithmic}

\usepackage{array}
\usepackage{makecell}

\usepackage{threeparttable}
\usepackage[table]{xcolor} 
\usepackage{booktabs}  
\usepackage{tabularx}  
\usepackage{amssymb}   
\begin{document}

\title{TransDOA: Calibrating Array Imperfections via Transformer-based Transfer Learning}

\author{Bo Zhou, Kaijie Xu, Yinghui Quan and Mengdao Xing, \IEEEmembership{Fellow, IEEE}
\thanks{This research was supported by the National Natural Science Foundation of China (Nos. 62101400, 72101075, 72171069 and 92367206), and in part by the Shaanxi Fundamental Science Research Project for Mathematics and Physics under Grant 22JSQ032. {\itshape (Corresponding author: Kaijie Xu)}}
\thanks{Bo Zhou, Kaijie Xu and Yinghui Quan are with the School of Information Mechanics and Sensing Engineering, Xidian University, Xi’an 710071, China (e-mail:23021211623@stu.xidian.edu.cn; kjxu@xidian.edu.cn; yhquan@mail.xidian.edu.cn). Mengdao Xing is with the National Laboratory of Radar Signal Processing, Xidian University, Xi'an 710071, China (email: xmd@xidian.edu.cn). }}

\markboth{Journal of \LaTeX\ Class Files,~Vol.~14, No.~8, August~2021}%
{Shell \MakeLowercase{\textit{et al.}}: A Sample Article Using IEEEtran.cls for IEEE Journals}


\maketitle

\begin{abstract}
	In practical scenarios, processes such as sensor design, manufacturing, and installation will introduce certain errors. Furthermore, mutual interference occurs when the sensors receive signals. These defects in array systems are referred to as array imperfections, which can significantly degrade the performance of Direction of Arrival (DOA) estimation. In this study, we propose a deep-learning based transfer learning approach, which effectively mitigates the degradation of deep-learning based DOA estimation performance caused by array imperfections.
	
	In the proposed approach, we highlight three major contributions. First, we propose a Transformer based method for DOA estimation, which achieves excellent performance in scenarios with low signal-to-noise ratios (SNR) and limited snapshots. Second, we introduce a transfer learning framework that extends deep learning models from ideal simulation scenarios to complex real-world scenarios with array imperfections. By leveraging prior knowledge from ideal simulation data, the proposed transfer learning framework significantly improves deep learning-based DOA estimation performance in the presence of array imperfections, without the need for extensive real-world data. Finally, we incorporate visualization and evaluation metrics to assess the performance of DOA estimation algorithms, which allow for a more thorough evaluation of algorithms and further validate the proposed method. Our code can be accessed at {\url{https://github.com/zzb-nice/DOA_est_Master}}.
\end{abstract}


\section{INTRODUCTION}
\begin{figure}[t]
	\begin{center}
		\centering
		\includegraphics[width=\linewidth]{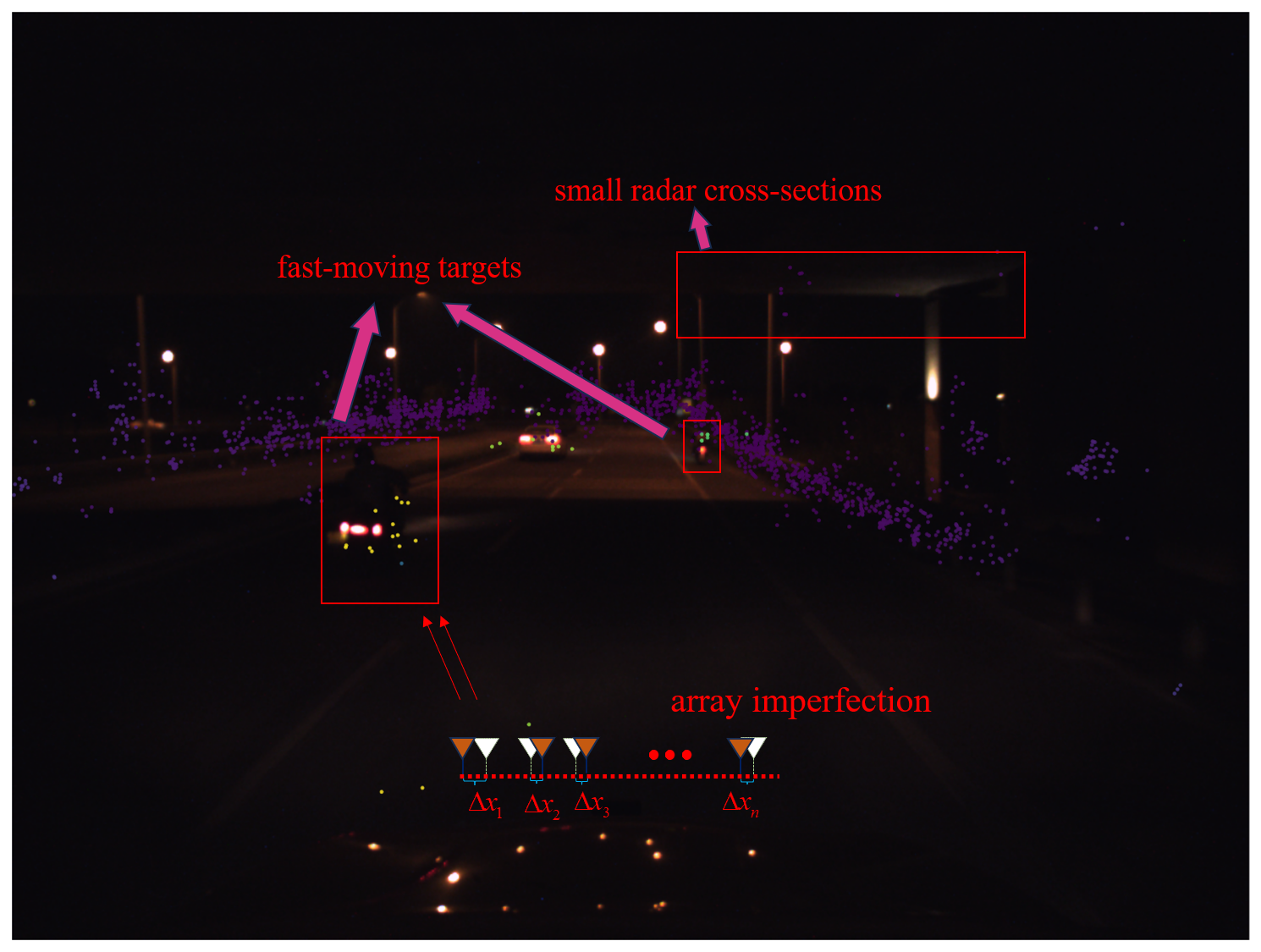}
		\caption{Schematic illustration of environmental perception using front-mounted mmWave radar in a night-time autonomous driving scenario. The radar point clouds are projected onto the camera image for visualization.}
		
		\label{autonomous_scenario}		
	\end{center}
\end{figure}

Direction of Arrival (DOA) estimation  plays a pivotal role across diverse engineering disciplines, including wireless communications, radio astronomy, radar, and sonar \cite{wangOverviewEnhancedMassive2019a,xuRobustDOAEstimation2023,Ben-DavidParametric2008,saucanCPHDDOATrackingMultiple2016}, where precise signal localization is critical. Recently, the proliferation of accessible wireless sensing technology has expanded the scope of DOA algorithms to emerging commercial domains, such as autonomous driving and unmanned aerial systems (UAS)  \cite{zhangSuperResolutionDOA2020,wanDeepLearningBased2021b,fan4DmmWaveradar2024,akterRFDOANetEfficientConvNet2021a}.

Rapid advancements in engineering have imposed increasingly stringent requirements on DOA estimation algorithms. For instance, autonomous driving necessitates the tracking of fast-moving targets with small radar cross-sections \cite{zhangSuperResolutionDOA2020}, while non-cooperative target localization demands both high speed and precision \cite{songCooperativeSensingNonCooperative2024}. In practice, these tasks are significantly complicated by the coexistence of challenging signal environments (e.g., low SNR and limited snapshots) and inevitable hardware impairments.

Fig. 1 illustrates the impact of signal-level and hardware-level challenges \cite{zheng2022tj4dradset}. In this real-world autonomous driving scenario, their interplay leads to severe performance degradation, manifested as significant point cloud sparsity and missed detections (e.g., the sparsely represented vehicle in the red box). However, existing algorithms often fail to address these coupled constraints simultaneously. Therefore, there is a compelling need for a new algorithm that delivers robust performance against array mismatches while maintaining high accuracy in low SNR and limited snapshot regimes.

Existing DOA estimation methods can be broadly categorized into model-driven and data-driven approaches. Among them, model-driven methods have been more extensively investigated due to their well-established theoretical foundations. Based on different estimation strategies, model-driven approaches can be further classified into several primary categories, including beamforming methods \cite{nanyanwangNewDOAEstimation2006a}, subspace-based methods \cite{haardtUnitaryESPRITHow1995,schmidtMultipleEmitterLocation1986b,yanRealvaluedRootMUSICDOA2018a}, maximum likelihood methods \cite{qiuMaximumLikelihoodMethod2021a,mellerDoAEstimationRotating2020a}, and compressed sensing (sparsity-inducing) methods \cite{stoicaSPICESparseCovarianceBased2011,malioutovSparseSignalReconstruction2005c,zaiyangEnhancingSparsityResolution2016b,liuEfficientMaximumLikelihood2012}.

A leading and well-respected family of model-driven approaches is the subspace-based methods, which achieve satisfactory performance in scenarios with high SNR and large snapshots.  However, their performance relies heavily on the asymptotic accuracy of the covariance matrix, which drastically degrades under low SNR and limited snapshot conditions. Moreover, these methods struggle with color noise, and require the estimated number of signal sources to be smaller than the number of array elements, which limits their applicability.

Even if these signal-level limitations could be overcome, a more fundamental, physical challenge remains: the divergence between the idealized model and the physical hardware. Most standard model-driven methods typically rely on an ideal array manifold. However, The existence of array position biases, amplitude and phase inconsistencies, mutual coupling \cite{allen2005adaptive}, and other array imperfections can severely distort the received data. In such cases, the mismatch between the theoretical model and the physical array often leads to significant performance degradation or even the complete failure of DOA estimation.

To mitigate these hardware impairments, various active calibration \cite{liu2DDOAEstimation2021,boonchongngSensorarrayCalibrationUsing1996,ngActiveArraySensor1995} and self-calibration \cite{liuUnifiedFrameworkSparse2013,zhangDOAPhaseError2020a,weissiEigenstructureMethodsDirection,liuEigenstructureMethodEstimating2011a,hanCalibratingNestedSensor2015,selloneNovelOnlineMutual2007,yeResiliencyMUSICDirection2008,liuDOAEstimationUniform2009} techniques have been developed. Active calibration methods, while affording high accuracy through the utilization of predefined auxiliary sources, necessitate significant resource expenditure and introduce operational discontinuity. In contrast, self-calibration methods offer a more practical solution by estimating array parameters and DOAs simultaneously without auxiliary sources. However, these methods often rely heavily on initialization and struggle to handle more severe array imperfections. As noted in \cite{liuUnifiedFrameworkSparse2013}, most existing self-calibration methods are ineffective in practical scenarios with low SNR, limited snapshots, and spatially proximate sources——precisely the challenges that many DOA estimation algorithms face.

Moreover, existing calibration methods often lack versatility. Due to the complexity of modeling combined imperfections, most algorithms focus on correcting only a single type of array imperfections, such as array position biases \cite{liu2DDOAEstimation2021,ngActiveArraySensor1995}, amplitude-phase inconsistencies \cite{zhangDOAPhaseError2020a,weissiEigenstructureMethodsDirection,liuEigenstructureMethodEstimating2011a,hanCalibratingNestedSensor2015}, or mutual coupling \cite{selloneNovelOnlineMutual2007,yeResiliencyMUSICDirection2008,liuDOAEstimationUniform2009}. Only a few studies have explored to address multiple simultaneous imperfections \cite{liuUnifiedFrameworkSparse2013,boonchongngSensorarrayCalibrationUsing1996}, and these often rely on specific array geometries (e.g., ULA/UCA) or other restrictive assumptions \cite{liuUnifiedFrameworkSparse2013}. Consequently, developing a robust framework capable of handling multiple array imperfections simultaneously, especially under low SNR and limited snapshot regimes, remains an urgent and open research problem.

In recent years, the rapid advancement of computational hardware \cite{lecunDeepLearning2015} has significantly accelerated the development of deep learning technologies. Empirical studies have demonstrated the superior capability of deep learning-based methods to approach the Cramér-Rao Lower Bound (CRLB) in challenging signal regimes characterized by low SNR and limited snapshots \cite{papageorgiouDeepNetworksDirectionofArrival2021,liuDirectionofArrivalEstimationBased2018}. As a result, data-driven methods are considered promising tools for achieving DOA estimation in practical scenarios.
Additionally, recent studies have concentrated on combining traditional DOA estimation methodologies with deep learning-based architectures to realize further advancements in performance. In 2022, the method proposed in \cite{wuGridlessDOAEstimation2022} integrated subspace-based methods with deep neural networks. Specifically, a CNN was employed to reconstruct a noise-free covariance matrix, after which a classical subspace-based approach was applied for DOA estimation. Similarly, the approaches in \cite{xiangAngleSeparationLearning2021,xuDeepNeuralNetworks2023} combined deep learning with compressed sensing algorithms. In these works, a DNN is utilized to the extend the covariance matrix to a higher-dimensional space or to compute prior knowledge for the compressed sensing algorithm. Then a sparse reconstruction methods is applied to estimate the incident angles. Furthermore, in 2024, Shmuel et al. \cite{shmuelSubspaceNetDeepLearningAided2024} introduced SubspaceNet, which directly treated the Root MUSIC algorithm as part of the network. Unlike previous work, SubspaceNet utilized the differentiability of polynomial equation root-finding and matrix decomposition, which allowed for the backpropagation of loss from the Root MUSIC algorithm.

Although these deep learning architectures have proven highly effective in challenging signal environments, the potential of utilizing deep learning techniques to mitigate physical array imperfections remains significantly unexplored in existing literature. While Liu et al. \cite{liuDirectionofArrivalEstimationBased2018} suggested that deep learning algorithms exhibit good adaptability to array imperfections, this adaptability does not extend to all operating conditions. As will be empirically demonstrated in Fig. \ref{array_imperfections} and Fig. \ref{pre_result_snr_5_rho_1}, deep learning models exhibit a pronounced sensitivity to array imperfections in challenging signal environments. Therefore, developing a robust deep learning-based approach capable of handling the simultaneous coexistence of adverse signal conditions and inevitable hardware impairments remains a critical and unresolved research challenge.

As an emerging deep learning technique, transfer learning algorithm \cite{weissSurveyTransferLearning2016} is capable of leveraging knowledge acquired from a source task to enhance the performance of a distinct target task. This characteristic holds the potential to enhance algorithmic resilience in practical scenarios, particularly where challenging signal environments and inevitable hardware impairments coexist. Nonetheless, there is currently a lack of research on the application of transfer learning methods in the field of DOA estimation. To the best of our knowledge, only a few papers \cite{wuGridlessDOAEstimation2024,zhangIntelligentDOAEstimation2022,labbafRobustDoAEstimation2023} have introduced transfer learning methods into DOA estimation. In \cite{wuGridlessDOAEstimation2024}, Wu et al. migrated a network designed for DOA estimation with multiple targets to one for fewer targets, with the primary goal of reducing the computational cost of training. Similarly, in \cite{labbafRobustDoAEstimation2023}, Labbaf et al. migrated a network designed for single-target DOA estimation to multiple targets. In \cite{zhangIntelligentDOAEstimation2022}, Zhang et al. applied transfer learning to enhance the network's adaptability to array imperfections, but they only fine-tuned the network from ideal data to array imperfections-affected data, without incorporating more advanced transfer learning techniques.

In this paper, we propose a Transformer-based transfer learning framework for DOA estimation, which aims to mitigate the performance degradation of data-driven approaches in practical scenarios with various array imperfections. In our approach, the data is divided into source and target domains. The source domain data consists of ideal data without array imperfections, while the target domain data represents practical scenarios containing array imperfections. Firstly, we propose a DOA estimation model, termed TransDOA, which is trained initially on the source domain data. A transfer learning method is then employed to align the features between the source and target domains, thereby significantly enhancing TransDOA's ability to adapt to array imperfections and improving performance in scenarios with low SNR, limited snapshots, and multiple array imperfections. Moreover, in a variety of scenarios, extensive experiments have been conducted to validate the effectiveness of the proposed approach. The contributions of our work can be outlined as follows:
\begin{enumerate}
	\item We propose a novel DOA estimation model, TransDOA. This model demonstrates exceptional performance under challenging conditions, such as low SNR and limited snapshot scenarios.
	\item Due to the presence of array imperfections, the data distributions of the source and target domains differ significantly, leading to substantial performance degradation when the model is deployed in practical scenarios. To address this issue, we introduce a transfer learning algorithm to align the features between the source and target domains, enhancing TransDOA's performance in practical scenarios.
	\item Via extensive simulations, we compare the proposed method with existing approaches across multiple evaluation metrics and demonstrate the superiority of our method in terms of DOA estimation accuracy and robustness.
\end{enumerate}

This paper is organized as follows. Section II formulates the problem. Section III presents the design of a network for DOA estimation. Section IV introduces a transfer learning scheme for the developed model. Experimental results are given in Section V. Finally, Section VI concludes the paper.

\section{Problem Formulation}

\begin{figure*}[t]
	\begin{center}
		\centering
		\includegraphics[width=\linewidth]{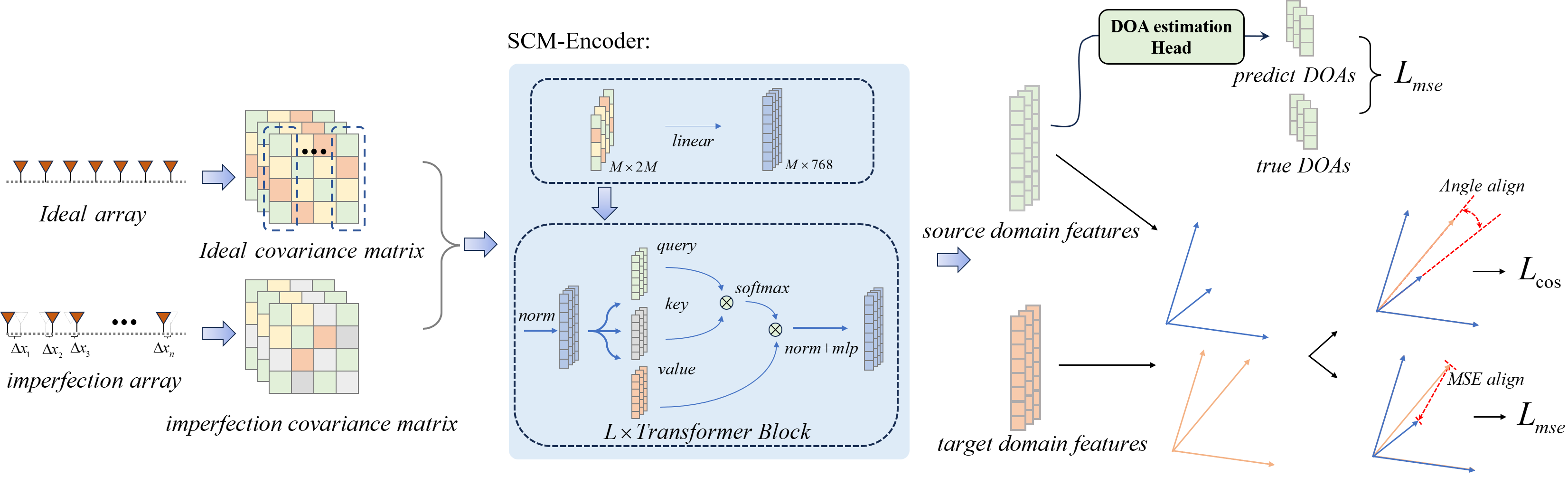}
		\caption{The framework of TransDOA and transfer learning approach proposed for array imperfections.}
		\label{total_framework}		
	\end{center}
\end{figure*}

In this work, two models of array receiving signals are considered. The first is an ideal array signal model, while the second incorporates array imperfections to reflect real-world conditions.
\subsection{Ideal Signal Model}  \label{section:2.1}
Firstly, the ideal array reception model is constructed. Consider a Uniform Linear Array (ULA) or Uniform Circular Array (UCA) consisting of $M$ antennas, receiving $K$ narrowband non-coherent far-field signals impinging from different DOAs. The array received signal model can be expressed in the following manner:

\begin{equation}
	\begin{aligned}
		\boldsymbol{y}(t) & =\sum_{k=1}^K \boldsymbol{a}\left(\boldsymbol{p},\theta_k\right) s_k(t)+\boldsymbol{n}(t) \\
		& =\boldsymbol{A}(\boldsymbol{p},\boldsymbol{\theta}) \boldsymbol{s}(t)+\boldsymbol{n}(t), t=1, \ldots, T .
	\end{aligned}
\end{equation}
where $\boldsymbol{\theta}=\left[\theta_1, \ldots, \theta_K\right]$ represents the directions of arrival, $\boldsymbol{A}(\boldsymbol{p},\boldsymbol{\theta})=\left[\boldsymbol{a}\left(\boldsymbol{p},\theta_1\right), \ldots, \boldsymbol{a}\left(\boldsymbol{p},\theta_K\right)\right]$ denotes the $M\times K$ array manifold matrix, $t=1,\ldots ,T$ is the time index, and $T$ is the number of received snapshots. $\boldsymbol{s}(t)={{[{s_{1}}(t),\ldots,{s_{K}}(t)]}^{T}}$ and $\boldsymbol{n}(t)$ denote the transmitted signal and the noise vector at time index $t$, respectively. The steering vectors for ULA and UCA can be expressed in the following form: 

\begin{equation}
		\boldsymbol{a}(\boldsymbol{p},\theta_k)=[1, e^{-j \frac{2 \pi d}{\lambda} \sin (\theta_k)}, \ldots, e^{-j \frac{2 \pi d}{\lambda} (M-1) \sin (\theta_k)}]^T
\end{equation}
\begin{equation}
	\begin{aligned}
		\boldsymbol{a}\left(\boldsymbol{p},\theta_k,\phi_k\right) = & \left[ e^{-j\frac{2 \pi R}{\lambda}\cos(\phi_1 - \theta_k)\sin(\phi_k)} , \dots, \right.\\ 
		& \left. e^{-j\frac{2 \pi R}{\lambda}\cos(\phi_M - \theta_k)\sin(\phi_k)}  \right]^T
	\end{aligned}
\end{equation}
where $\theta_k$ and $\phi_k$ denote the DOA of the $k$-th signal. $\lambda =c/f$ denotes the wavelength of the incident signals, $d$ is the distance between the array elements in ULA and $R$ is the radius of the UCA. $\phi_m = \frac{2\pi m}{M}$ denotes the azimuth of the $m$th sensor of the UCA. Based on the received signal discussed above, the sampled covariance matrix (SCM) is computed as follows:
\begin{equation}
	\widetilde{\boldsymbol{R}}_y=\frac{1}{T} \sum_{t=1}^T \boldsymbol{y}(t) \boldsymbol{y}^{\mathrm{H}}(t)
\end{equation}
$\widetilde{\boldsymbol{R}}_y$ is an unbiased estimate of ideal covariance matrix calculated by the received signal vector. Most existing DOA estimation algorithms focus on deriving the source DOAs from the sampled SCM $\widetilde{\boldsymbol{R}}_y$, which has been proven to be an effective approach.

\subsection{Signal Model with Array Imperfection}
In practical applications, the array signal reception model is often deviates from its theoretical counterpart. Such non-ideal conditions typically arise from three primary sources: position biases introduced during array element arrangement; amplitude and phase inconsistencies caused by design and manufacturing defects; and mutual coupling effects induced during the signal reception process. In \cite{liuDirectionofArrivalEstimationBased2018}, Liu et al. established a simplified model to demonstrate how these three array imperfections affect the signal model in a ULA, introducing a hyperparameter $\rho \in [0,1]$ to succinctly control the intensity of these effects.
\begin{equation}
	\scalebox{0.95}{$\begin{aligned}
			\boldsymbol{e}_{\text{pos}} &= \rho \times \left[ 0, -0.2, \dots, -0.2, 0.2, 0.2, \dots, 0.2 \right]^T \times d \\
			\boldsymbol{e}_{\text{gain}} &= \rho \times \left[ 0, 0.2, \dots, 0.2, -0.2, \dots, -0.2 \right]^T \\
			\boldsymbol{e}_{\text{phase}} &= \rho \times \left[ 0, -30^\circ, \dots, -30^\circ, 30^\circ, \dots, 30^\circ \right]^T \\
			\boldsymbol{e}_{\text{mc}} &= \rho \times \left[ \gamma^0, \gamma^1, \dots, \gamma^{M-1} \right]^T
		\end{aligned}$}
\end{equation}
where \( \gamma = 0.3e^{j60^\circ} \) is the mutual coupling coefficient between adjacent sensors. The vectors \( \boldsymbol{e}_{\text{pos}} \), \( \boldsymbol{e}_{\text{gain}} \) and \( \boldsymbol{e}_{\text{phase}} \) denote the element positional, amplitude, and phase biases, respectively. Specifically, \( \boldsymbol{e}_{\text{mc}} \) models the magnitude of the mutual coupling effect in a ULA, where the strength of the mutual coupling is proportional to the square of the distance between elements. The overall mutual coupling across the entire array is represented by the matrix \( \boldsymbol{E}_{\text{mc}} \), which is defined as a Toeplitz matrix parameterized by the vector \( \boldsymbol{e}_{\text{mc}} \) \cite{friedlanderDirectionFindingPresence1991a}:
\begin{equation}
	\small
	\boldsymbol{E}_{\text{mc}} =  
	\begin{bmatrix}  
		\gamma^0 & \gamma^1 & \cdots & \gamma^i & \gamma^{i+1} & \cdots & \gamma^{M-1} \\
		\gamma^1 & \gamma^0 & \gamma^1 & \cdots & \gamma^i & \cdots & \vdots \\
		\vdots & \gamma^1 & \gamma^0 & \ddots & \cdots & \ddots & \vdots \\
		\gamma^i & \cdots & \ddots & \ddots & \ddots & \ddots & \gamma^{i} \\
		\gamma^{i+1} & \ddots & \cdots & \ddots & \ddots & \ddots & \vdots \\
		\vdots & \ddots & \ddots & \cdots & \ddots & \gamma^0 & \gamma^1 \\
		\gamma^{M-1} & \cdots & \gamma^{i+1} & \gamma^i & \cdots & \gamma^1 & \gamma^0  
	\end{bmatrix}
\end{equation}
Similarly, for a UCA, the \( \boldsymbol{e}_{\text{mc}} \) is exponentially related to the inter-element spacing. The position bias \( \boldsymbol{e}_{\text{pos}} \) can be extended to both the \(x\)- and \(y\)-axes, while \( \boldsymbol{e}_{\text{gain}} \) and \( \boldsymbol{e}_{\text{phase}} \) remain unchanged. The steering vector affected by all three array imperfections can be expressed as:

\begin{equation}
	\scalebox{0.9}{$\begin{aligned}
			\boldsymbol{a}(\theta,\boldsymbol{e}) &= \left(\boldsymbol{I}_{M} + \delta_{mc} \boldsymbol{E}_{mc}\right) \times \left(\boldsymbol{I}_{M} + \operatorname{Diag}\left(\delta_{gain} \boldsymbol{e}_{gain}\right)\right) \\
			&\times \operatorname{Diag}\left(\exp\left(j\delta_{\text{phase}} \boldsymbol{e}_{\text{phase}}\right)\right) \times \boldsymbol{a}\left(\theta,\boldsymbol{p}+\delta_{pos} \boldsymbol{e}_{pos}\right)
		\end{aligned}$}
	\label{eq6}
\end{equation}
where \(\delta(\cdot)\) indicates the presence of specific array imperfections. \(\boldsymbol{I}_M\) represents the \(M \times M\) identity matrix, while \(\operatorname{Diag}(\cdot)\) denotes a diagonal matrix formed from a given vector. And \(\boldsymbol{a}\left(\theta,\boldsymbol{p}+ \boldsymbol{e}_{pos}\right)\) refers to the array steering vector incorporating the position biases:
\begin{equation}
	\begin{aligned}
		\boldsymbol{a}\left(\theta_k, \boldsymbol{p} +  \boldsymbol{e}_{pos}\right) =& [ e^{-j \frac{2 \pi}{\lambda} (0d+e_{\text{pos}}^0) \sin(\theta_k)}, e^{-j \frac{2 \pi}{\lambda} (1d+e_{\text{pos}}^1) \sin(\theta_k)},\\ &\ldots, e^{-j \frac{2 \pi}{\lambda} ((N-1)d+e_{\text{pos}}^{N-1}) \sin(\theta_k)}  ]^{T}
	\end{aligned}
	\raisetag{40pt}
\end{equation}
where \( e_{\text{pos}}^i \) represents the position biases of the \( i \)th antenna. Therefore, the final result \( \boldsymbol{a}(\theta,\boldsymbol{e}) \) in Equation \eqref{eq6} represents the steering vector influenced by the three types of array imperfections. Therefore, the imperfection steering matrix \(\boldsymbol{A}\) can be expressed as: 
\begin{equation}
	\boldsymbol{A}(\boldsymbol{\theta}, \boldsymbol{e})=\left[ \boldsymbol{a}(\theta_1,\boldsymbol{e}),\boldsymbol{a}(\theta_2,\boldsymbol{e}),\ldots,\boldsymbol{a}(\theta_K,\boldsymbol{e}) \right]
\end{equation}
By substituting \( \boldsymbol{A}(\boldsymbol{\theta}, \boldsymbol{e}) \) for the steering matrix \( \boldsymbol{A}(\boldsymbol{p}, \boldsymbol{\theta}) \) introduced in Section \ref{section:2.1}, a more accurate simulation of the array signal reception model in practical scenarios can be achieved.
\section{NETWORKS FOR DOA ESTIMATION}
\begin{figure*}[t]
	\begin{center}
		\centering
		\includegraphics[width=0.9\linewidth]{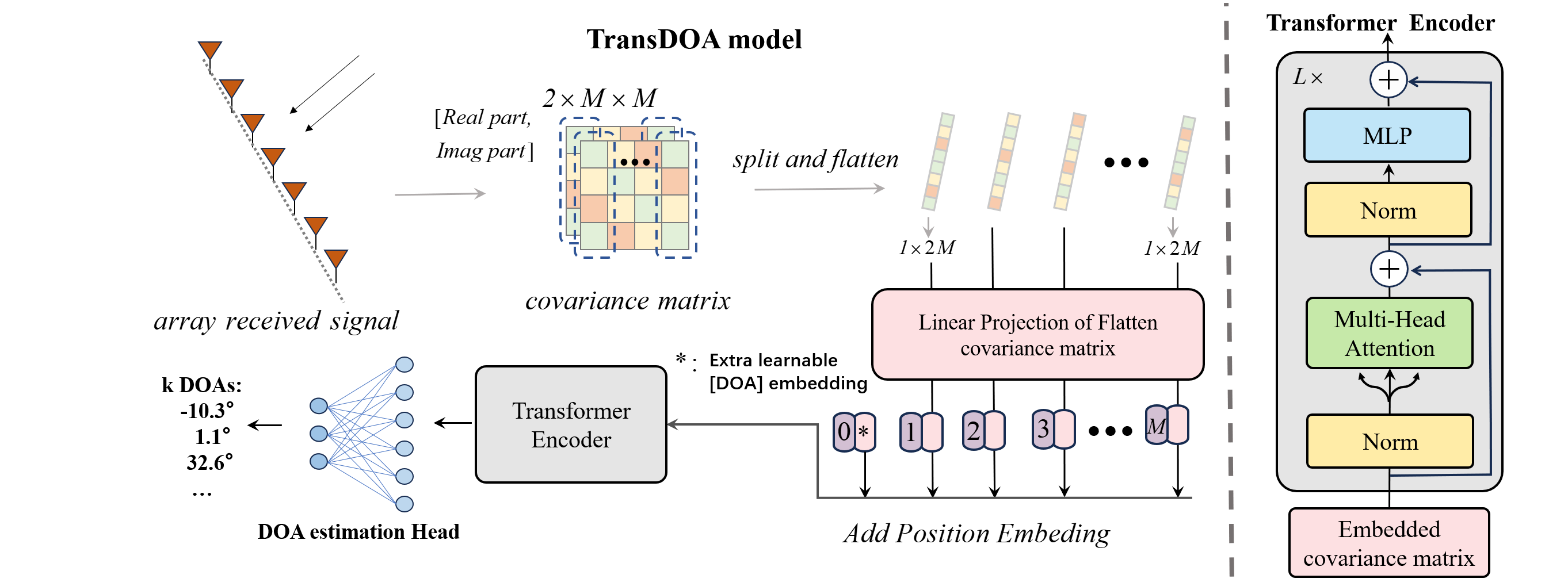}
		\caption{The overall structure of TansDOA and the processing procedure of input covariance matrix.}
		\label{transdoa}		
	\end{center}
\end{figure*}
To capture global dependencies among array elements, we propose a Transformer-based model named TransDOA. The overall framework comprising the TransDOA architecture and the transfer learning strategy is illustrated in Fig. \ref{total_framework}. Specifically, the detailed structure of the TransDOA is further presented in Fig. \ref{transdoa}.

\subsection{Covariance Matrix Embedding}

To accommodate the Transformer's sequence-based architecture for DOA estimation, the SCM $\widetilde{\boldsymbol{R}}_y$ is partitioned and mapped into a sequence of vectors. Specifically, $\widetilde{\boldsymbol{R}}_y$ is split column-wise into $M$ blocks, each flattened and transformed into a $D$-dimensional embedding $\boldsymbol{x}_i$ via a linear layer $\boldsymbol{E}$, as formulated in (\ref{eq7})--(\ref{eq9}). A learnable DOA token is prepended to the input sequence to aggregate global features for the final estimation task (\ref{eq10}).
\begin{align}
	\widetilde{\boldsymbol{R}}_y &= \left[\boldsymbol{r}_1, \boldsymbol{r}_2, \boldsymbol{r}_3, \ldots, \boldsymbol{r}_M\right],\;\small{ \widetilde{\boldsymbol{R}}_y \in \mathbb{C}^{M \times M},r_i \in \mathbb{C}^{M}}
	\label{eq7}\\
	\widetilde{\boldsymbol{r}}_i &= \left[\text{real}(\boldsymbol{r}_i); \text{imag}(\boldsymbol{r}_i)\right], \quad \widetilde{\boldsymbol{r}}_i \in \mathbb{R}^{2 M} \label{eq8}\\
	\boldsymbol{x}_i &= \boldsymbol{E} \widetilde{\boldsymbol{r}}_i, \quad \boldsymbol{E} \in \mathbb{R}^{D \times 2 M} \label{eq9}\\
	\boldsymbol{X}_{\text {input }} &= \left[doa\;token, \boldsymbol{x}_1, \boldsymbol{x}_2, \ldots, \boldsymbol{x}_M\right]  \label{eq10}
\end{align}

Since the self-attention mechanism is permutation-invariant, learnable positional embeddings $\boldsymbol{E}_{pos}$ are added to the input representations to preserve the spatial structure of the array signal:
\begin{equation}
	\begin{aligned}
		\boldsymbol{z}_{\text {input }} & =\left[\boldsymbol{x}_0+\boldsymbol{e}_0, \boldsymbol{x}_1+\boldsymbol{e}_1, \cdots, \boldsymbol{x}_M+\boldsymbol{e}_M\right] \\
		& =\boldsymbol{X}_{\text {input }}+\boldsymbol{E}_{\text {pos }}, \quad \boldsymbol{E}_{\text {pos }} \in \mathbb{R}^{D \times (M+1)}
	\end{aligned}
\end{equation}

Further details regarding the SCM Embedding process and the comprehensive configuration of the TransDOA model are presented in Fig. \ref{transdoa}.

\subsection{Transformer Encoder}
The Multi-Head Self-Attention (MHSA) mechanism serves as the core component of the Transformer Encoder, designed to capture complex global dependencies within the input sequence.

Given an input \( \boldsymbol{Z} \in \mathbb{R}^{D \times (M+1)} \), where \( D \) is the feature dimension and \( M+1 \) is the sequence length, the MHSA operates are defined as follow.

\textbf{Linear Projections:} The input is linearly projected into \textit{queries} \( \boldsymbol{Q} \), \textit{keys} \( \boldsymbol{K} \), and \textit{values} \( \boldsymbol{V} \) for each attention head \( i \):
\begin{align}
	\boldsymbol{Q}_i = \boldsymbol{W}_i^Q \boldsymbol{Z}, \quad \boldsymbol{K}_i = \boldsymbol{W}_i^K \boldsymbol{Z}, \quad \boldsymbol{V}_i = \boldsymbol{W}_i^V \boldsymbol{Z}, \quad i = 1, \dots, h
\end{align}
where \( \boldsymbol{W}_i^Q, \boldsymbol{W}_i^K, \boldsymbol{W}_i^V \in \mathbb{R}^{d_k \times D} \) are model parameters, and \( d_k = D / h \) is the dimension of each head.

\textbf{Scaled Dot-Product Attention:} Each head computes attention as:
\begin{align}
	\text{head}_i = \text{softmax}\left( \frac{\boldsymbol{Q}_i \boldsymbol{K}_i^T}{\sqrt{d_k}} \right)\boldsymbol{V}_i
\end{align}

\textbf{Concatenation and Output:} The outputs of all \( h \) heads are concatenated and projected to produce the final output:
\begin{align}
	\text{MHSA}(\boldsymbol{Z}) = \boldsymbol{W}^O \text{Concat}(\text{head}_1, \dots, \text{head}_h)
\end{align}
where \( W^O \in \mathbb{R}^{D \times h d_k} \) is the output projection matrix. As illustrated in Fig. \ref{transdoa}, the Transformer Encoder, consisting of layer normalization, MHSA and an MLP, enables the TransDOA model to effectively capture and integrate global information.

\subsection{DOA Estimation Head}
The DOA Estimation Head serves as the final regression module. The feature vector corresponding to the aggregated DOA token is extracted and projected via a linear layer to predict the target DOAs. To address the permutation ambiguity inherent in multi-source estimation, we employ the permutation invariant training loss:
\begin{equation}
	l(\boldsymbol{\widetilde{\boldsymbol{R}}}, \boldsymbol{\theta}; \boldsymbol{\psi}) = \min_{\boldsymbol{P} \in \mathcal{P}} \left( \frac{1}{K} \left\| \boldsymbol{\theta} - \boldsymbol{P} \hat{\boldsymbol{\theta}}(\widetilde{\boldsymbol{R}}; \boldsymbol{\psi}) \right\|^2 \right)^{\frac{1}{2}}
\end{equation}
where $\boldsymbol{\psi}$ denotes the model parameters and $\hat{\boldsymbol{\theta}}$ represents the estimation result, which includes $K$ DOAs. $\boldsymbol{\theta}$ corresponds to the true DOAs, and the set $\mathcal{P}$ encompasses all $K \times K$ permutation matrices, defined as binary matrices where each row and column contains exactly one non-zero element.

Furthermore, the proposed framework extends naturally to 2D (Two-Dimensional) DOA estimation. By extending the DOA Estimation Head to predict both azimuth $\theta$ and elevation $\phi$ angles, the joint loss function is formulated as:
\begin{equation}
	\begin{split}
		l(\boldsymbol{\widetilde{\boldsymbol{R}}}, \boldsymbol{\theta}, \boldsymbol{\phi}; \boldsymbol{\psi}) = & \min_{\boldsymbol{P} \in \mathcal{P}} \left( \frac{1}{2K} \left\| \boldsymbol{\theta} - \boldsymbol{P} \hat{\boldsymbol{\theta}}(\widetilde{\boldsymbol{R}}; \boldsymbol{\psi}) \right\|^2 \right. \\
		& \quad \left. + \frac{1}{2K} \left\| \boldsymbol{\phi} - \boldsymbol{P} \hat{\boldsymbol{\phi}}(\widetilde{\boldsymbol{R}}; \boldsymbol{\psi}) \right\|^2 \right)^{\frac{1}{2}}
	\end{split}
\end{equation}

\section{Transfer Learning Approach}
\begin{figure*}[t]
	\begin{center}
		\centering
		\includegraphics[width=0.9\linewidth]{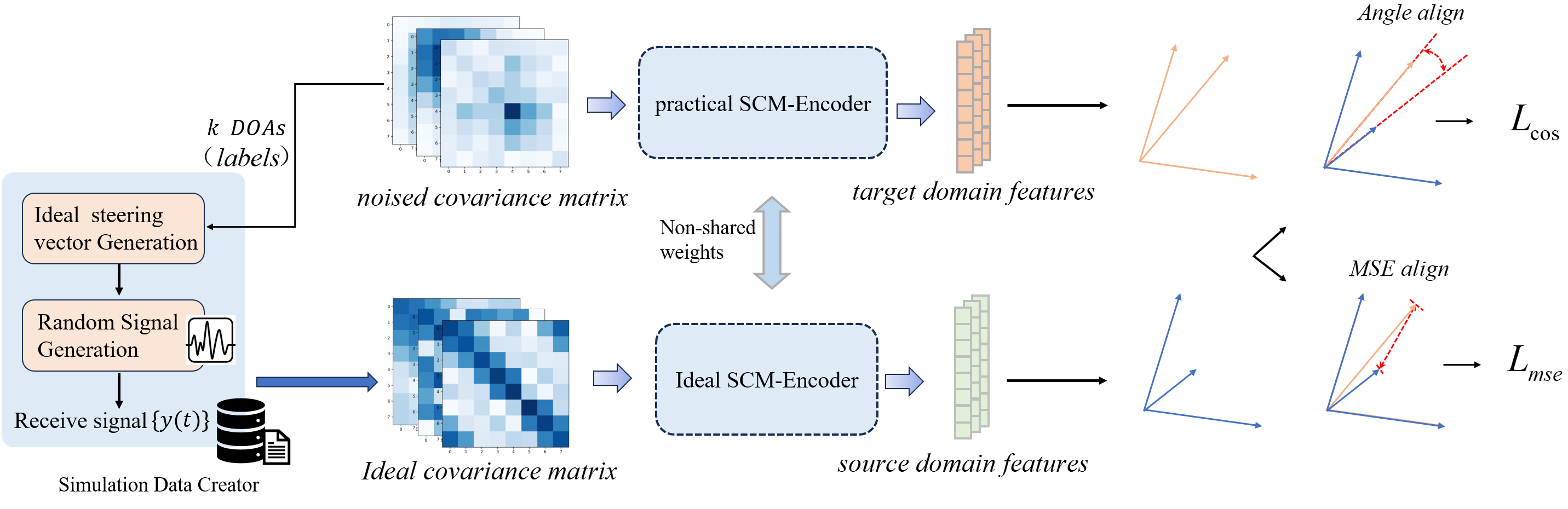}
		\caption{The framework of transfer learning approach proposed for array imperfections.}
		\label{transfer learning}		
	\end{center}
\end{figure*}
Deep learning models trained on large-scale simulated datasets often suffer from performance degradation in practical scenarios. This is primarily due to the \textbf{domain gap} caused by array imperfections.

To bridge this gap, we propose a supervised transfer learning framework based on \textbf{feature alignment}. As illustrated in Fig. \ref{transfer learning}, our approach aims to train a target domain feature extractor that maps distorted SCM into a domain-invariant feature space aligned with the ideal array manifold.

\subsection{Domain Characterization}
In our approach, two distinct scenarios are considered. The first is the ideal reception scenario, a standard paradigm in deep learning research. In this scenario, large-scale training datasets can be generated via simulation at low cost, enabling effective model training and superior performance.

The second scenario pertains to practical applications characterized by array imperfections. In contrast to the ideal case, data acquisition here relies on physical measurements, inevitably resulting in high costs and data scarcity. Based on these characteristics, we formally define the respective domains as follows:
\begin{itemize}
	\item \textbf{Source Domain ($\mathcal{D}_s$):} Consists of abundant ideal SCMs generated via simulation. A model trained here captures the theoretical array manifold but fails on imperfect data.
	\item \textbf{Target Domain ($\mathcal{D}_t$):} Consists of limited labeled data from the practical system with array imperfections.
\end{itemize}

How to effectively leverage the prior knowledge acquired from a model trained on a substantial amount of simulation data to enhance model's performance in a data-scarce or even data-absent practical setting becomes a valuable area for exploration.

\subsection{Transfer learning Framework}
To address the performance degradation caused by statistical discrepancies between the source and target domains, we propose a transfer learning framework, as illustrated in Fig. \ref{transfer learning}. By leveraging the pre-trained source parameters \(\boldsymbol{\psi}_{s}\), our objective is to train a target extractor that learns \textbf{domain-invariant features}, thereby ensuring robustness against array imperfections.

Specifically, we remove the DOA estimation Head from the TransDOA model and utilize the backbone as a feature extractor, denoted as $\mathcal{F}$. The feature representations for the source and target domains are obtained as:
\begin{equation}
	\begin{aligned}
		\boldsymbol{z}_{s,i} &= \mathcal{F}_{s}(\widetilde{\boldsymbol{R}}_{s}^{(i)}; \boldsymbol{\psi}_{s}) \\
		\boldsymbol{z}_{t,i} &= \mathcal{F}_{t}(\widetilde{\boldsymbol{R}}_{t}^{(i)}; \boldsymbol{\psi}_{t}) \\
	\end{aligned}
\end{equation}
where \( \mathcal{F}_{s} \) and \( \mathcal{F}_{t} \) are the feature extractors for the source and target domains, parameterized by $\boldsymbol{\psi}_{s}$ and $\boldsymbol{\psi}_{t}$, respectively. The vectors \( \boldsymbol{z}_{s,i} \) and \( \boldsymbol{z}_{t,i} \) represent the feature extracted from the $i$-th pair of source and target inputs.

To ensure exact feature correspondence between the source and target domains, it is critical to construct a strictly paired dataset. A \textbf{Simulation Data Generator} is employed for this purpose. In practical scenarios, we collect a series of SCMs along with their corresponding DOA values. Subsequently, these true DOA values are fed into the Simulation Data Generator to generate corresponding ideal covariance matrices. This process establishes a one-to-one mapping, where each collected real-world sample is paired with a generated ideal counterpart, thereby forming the requisite paired dataset for effective knowledge transfer and model calibration.

Utilizing the constructed paired dataset, the transfer learning framework is illustrated in Fig. \ref{transfer learning}. The pre-trained source feature extractor $\mathcal{F}_{s}$ is frozen to serve as a stable reference. The target encoder $\mathcal{F}_{t}$ is then trained to align its extracted features \( \boldsymbol{z}_{t,i} \) with the reference features \( \boldsymbol{z}_{s,i} \). This alignment compels \( \mathcal{F}_{t} \) to suppress domain-specific distortions caused by array imperfections. To enforce such robust alignment, we propose a composite objective function consisting of two terms:

\noindent \textbf{1) MSE Loss:} The Mean Squared Error (MSE) is employed to minimize the Euclidean distance between the feature representations of the two domains:
\begin{equation}
	\mathcal{L}_{\text{mse}}(\boldsymbol{z}_i^{S \leftrightarrow T}) =  \|\boldsymbol{z}_{s,i} - \boldsymbol{z}_{t,i}\|^2
\end{equation}
where \( S \) and \( T \) represent the source and target domains, respectively. The notation \( \| \cdot \| \) denotes the \( l_2 \)-norm of a vector.

\noindent \textbf{2) Cosine Similarity Loss:} To further enforce structural consistency, we treat features as vectors in a high-dimensional space and align their geometric directions. The cosine loss is formulated as:
\begin{equation}  
	\mathcal{L}_{\cos}(\boldsymbol{z}_i^{S \leftrightarrow T}) = 1- \frac{\boldsymbol{z}_{s,i} \cdot \boldsymbol{z}_{t,i}}{\|\boldsymbol{z}_{s,i}\| \cdot \|\boldsymbol{z}_{t,i}\|}  
\end{equation} 

By minimizing this loss, we maximize the cosine similarity, ensuring the target features share the same angular direction as the source features. Finally, the overall objective function is a weighted sum of the aforementioned losses:
\begin{equation}
	\mathcal{L}_{\text{total}}(\boldsymbol{Z}^S, \boldsymbol{Z}^T) =  \alpha \mathcal{L}_{\text{cos}}(\boldsymbol{Z}^S, \boldsymbol{Z}^T) + \beta \mathcal{L}_{\text{mse}}(\boldsymbol{Z}^S, \boldsymbol{Z}^T)
	\label{total_loss}
\end{equation}
where \( \boldsymbol{Z}^S \) and \( \boldsymbol{Z}^T \) represent all feature vectors extracted from the source domain and target domain, respectively, while \( \alpha \) and \( \beta \) are hyper-parameters controlling the effect of the two loss functions. To provide a more precise depiction of the overall process, the transfer learning procedure of TransDOA using Adam is detailed as Algorithm \ref{alg:vitnet}.
\begin{algorithm}[H]
	\caption{Transfer learning Approach using Adam}
	\label{alg:vitnet}
	\begin{algorithmic}[1]
		\REQUIRE TransDOA weights $\boldsymbol{\psi}_s$ pretrained on the source domain, hyperparameter $\epsilon$.
		\REQUIRE Target domain Dataset $\mathcal{D}_{\text{target}}$, learning rate $\mu$, number of batches $B$, epochs number $e_{\max}$.
		\STATE Initialize TransDOA weights $\boldsymbol{\psi}_s$ for source domain model.
		\vspace{-1em}
		\STATE Initialize TransDOA weights $\boldsymbol{\psi}_t$ for target domain model.
		\vspace{-1em}
		\FOR{$\text{epoch} = 0, 1, \dots, e_{\max}$}
		\STATE Randomly divide $\mathcal{D}$ into $B$ batches $\{\mathcal{D}_b\}_{b=1}^B$.
		\FOR{$b = 1, 2, \dots, B$}
		\FOR{$(\widetilde{\boldsymbol{R}}_b^{(j)}, \boldsymbol{\theta}_b^{(j)}) \in \mathcal{D}_b$}
		\STATE A Simulation Data Generator creates ideal covariance matrices $\widetilde{\boldsymbol{R}}_b^{*}$ based on $\boldsymbol{\theta}_b^{(j)}$
		\STATE Compute representations $\boldsymbol{Z}^S(\widetilde{\boldsymbol{R}}_b^{(j)}, \boldsymbol{\psi}_s)$ for the source domain.
		\STATE Compute representations $\boldsymbol{Z}^T(\widetilde{\boldsymbol{R}}_b^{*}, \boldsymbol{\psi}_t)$ for the target domain.
		\ENDFOR
		\STATE Compute $\mathcal{L}_{\mathcal{D}}(\boldsymbol{Z}^S, \boldsymbol{Z}^T)$ through Equation \ref{total_loss}.
		\STATE Update weights via $\boldsymbol{\psi}_{t} \gets \boldsymbol{\psi}_{t} - \mu \nabla_\psi \mathcal{L}_{\mathcal{D}_b}(\boldsymbol{Z}^S, \boldsymbol{Z}^T)$.
		\ENDFOR
		\ENDFOR
		\RETURN $\boldsymbol{\psi}_{t}$
	\end{algorithmic}
\end{algorithm}

\section{SIMULATIONS AND ANALYSES}
In this section, we evaluate the performance of the proposed scheme through extensive simulations, comprehensively validating its effectiveness and robustness under varying critical factors, specifically \textbf{SNR, number of snapshots $T$, number of array elements $M$, number of incident signals $K$, and angular separation}.

Comprehensive comparisons are conducted with a wide range of classic and state-of-the-art model-driven methods, including high-resolution subspace methods such as \textit{MUSIC} and \textit{Unity-ESPRIT} algorithms, as well as the compressive sensing-based \textit{$\ell_1$-SVD} approach \cite{zaiyangEnhancingSparsityResolution2016b,eldarRobustRecoverySignals2009a}, which has strong performance in low SNR and small snapshots scenarios. Additionally, a series of deep learning methods and their variants that integrate traditional algorithms are considered, including the spatial spectrum estimation CNN algorithm (\textit{SPE-CNN})  \cite{papageorgiouDeepNetworksDirectionofArrival2021}, \textit{SubspaceNet} \cite{shmuelSubspaceNetDeepLearningAided2024}, \textit{ASL-2} \cite{xiangAngleSeparationLearning2021}, and \textit{Learning-SPICE} \cite{xuDeepNeuralNetworks2023}. \textit{SPE-CNN} is recognized as a well-established deep learning-based method and \textit{SubspaceNet} can be understood as a subspace algorithm implemented using deep learning networks. Furthermore, \textit{ASL-2} and \textit{Learning-SPICE}, which representing the latest algorithms that combine deep learning and compressive sensing, are expected to perform well in limited snapshots scenarios.

We begin with a detailed description of the experimental setup in Subsection \ref{section:5.1}. The performance of the proposed model under these setups is then presented in Subsection \ref{section:5.2}. Furthermore, in  Subsection \ref{section:5.3}, we validate the effectiveness of transfer learning strategies through simulations conducted in practical scenarios involving array imperfections.

\subsection{Experimental Setup} \label{section:5.1}

\begin{table}[htbp]
	\centering
	\caption{Simulation Scenarios and Parameter Configurations}
	\label{tab:scenarios_vars}
	\footnotesize
	\begin{threeparttable}
		
		\begin{tabularx}{\columnwidth}{c c X}
			\toprule
			\textbf{Scen.} & \textbf{Setting $(M, K)$} & \textbf{Configuration \& Test Conditions} \\
			\midrule
			
			\multirow{2}{*}{\textbf{1}} & \multirow{2}{*}{ULA $(8, 3)$} 
			& \textbf{DOA:} \textbf{Uniform}, \textbf{Equidistant} \\ 
			& & \textbf{Params [1]:} SNR $= -5$ dB, $T=10$ \\
			& & \textbf{Params [2]:} SNR $\in [-20, 5]$ dB, $T=10$ \newline \textbf{Params [3]:} $T \in [10, 100]$, SNR$=-10$ dB \\
			\cmidrule{1-3}
			
			\textbf{2} & ULA $(8, 7)$ 
			& \textbf{DOA:} \textbf{Uniform} \\
			& & \textbf{Params [1]:} SNR $\in [-20, 5]$ dB, $T=10$ \\
			\cmidrule{1-3}
			
			\multirow{2}{*}{\textbf{3}} & \multirow{2}{*}{ULA $(16, 3)$} 
			& \textbf{DOA:} \textbf{Uniform} \\
			& & \textbf{Params [1]:} SNR $\in [-20, 5]$ dB, $T=10$ \\
			\cmidrule{1-3}
			
			\textbf{4} & UCA $(12, 5)$ 
			& \textbf{DOA:} \textbf{Deterministic}, \textbf{Uniform} \\
			& & \textbf{Params [1]:} SNR $= -5$ dB, $T=50$ \\
			& & \textbf{Params [2]:} SNR $= 5$ dB, $T=10$ \\
			\bottomrule
		\end{tabularx}
		
		\begin{tablenotes}
			\footnotesize
			\item[] \textbf{Uniform:} DOAs are sampled from a uniform distribution, subject to a minimum angular separation of $3^\circ$ to avoid source clustering.
			\item[] \textbf{Equidistant:} DOAs are equally spaced across the entire angular sector.
			\item[] \textbf{Deterministic:} A predetermined set of specific DOAs.
		\end{tablenotes}
		
	\end{threeparttable}
\end{table}

\textit{1) Simulation Scenarios:} To facilitate a holistic assessment of the algorithm's performance, we conduct simulations across four distinct scenarios, as summarized in Table \ref{tab:scenarios_vars}.

\begin{figure*}[t]
	\begin{center}
		\centering
		\includegraphics[width=0.9\linewidth]{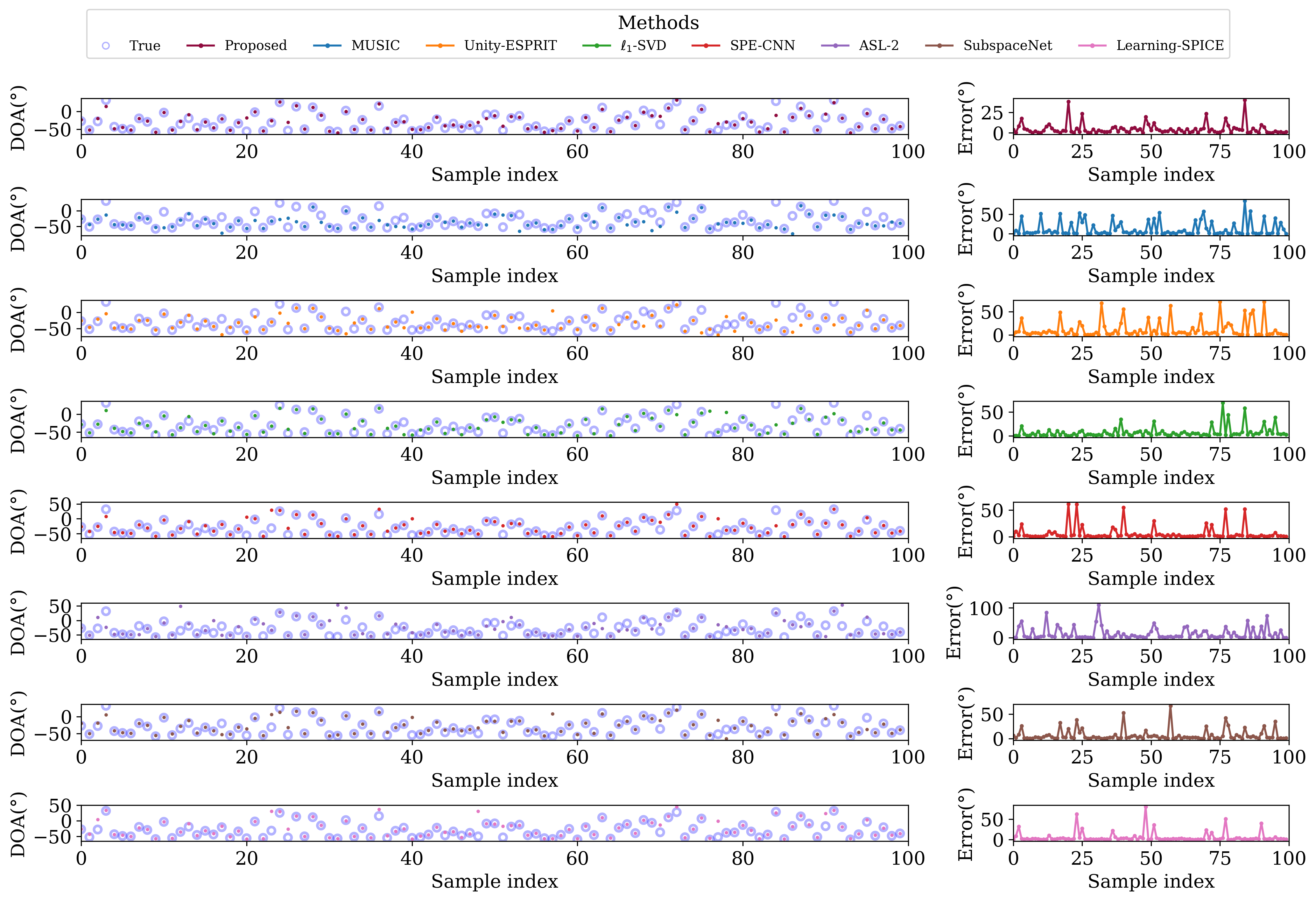}
		\caption{DOA estimation results under Scenarios 1 with $SNR=-5db$ and $snap=10$. Three incident signals are generate and we focus on the first one. The points represent the estimated DOA while the circles denote the true DOAs.}
		\label{result_snr_-5_snap_10}	
	\end{center}
\end{figure*}

\begin{table*}[htbp]
	\centering
	\caption{CDF-Based Statistical Evaluation results under Scenarios 1 with $SNR=-5$ and $snap=10$.}
	\label{comparison_arrows_moved}
	
	\resizebox{\textwidth}{!}{%
		\begin{tabular}{l c cc cc cc cc cc cc}
			\toprule
			\multirow{2}{*}{\textbf{Method}} & \multirow{2}{*}{\textbf{Miss Prob.} $\downarrow$} 
			& \multicolumn{2}{c}{\textbf{OSPA Error}} 
			& \multicolumn{2}{c}{\textbf{RMSE ($^\circ$)}} 
			& \multicolumn{2}{c}{\textbf{MAE ($^\circ$)}} 
			& \multicolumn{2}{c}{\textbf{Acc. (\%)}} 
			& \multicolumn{2}{c}{\textbf{\makecell{ECDF Quantile \\ (Raw)}}} 
			& \multicolumn{2}{c}{\textbf{\makecell{ECDF Quantile \\ (Matched)}}} \\
			
			\cmidrule(lr){3-4} 
			\cmidrule(lr){5-6} 
			\cmidrule(lr){7-8}   
			\cmidrule(lr){9-10}  
			\cmidrule(lr){11-12} 
			\cmidrule(lr){13-14}
			
			& & Linear $\downarrow$ & Square $\downarrow$
			& Raw $\downarrow$ & Matched $\downarrow$
			& Raw $\downarrow$ & Matched $\downarrow$
			& Raw $\uparrow$ & Matched $\uparrow$ 
			& 10\% $\downarrow$ & 90\% $\downarrow$ 
			& 10\% $\downarrow$ & 90\% $\downarrow$ \\
			\midrule
			
			MUSIC          & \cellcolor{black!10} 0.001 & 6.65 & 9.52 & 20.39 & 12.11 & 10.81 & 6.68 & 74.3 & 80.9 & 0.43 & 37.31 & 0.38 & 30.00 \\
			Unity-ESPRIT   & 0.000 & 6.27 & 8.46 & 17.09 & 10.52 & 8.92  & 6.29 & 78.8 & 82.8 & 0.50 & 26.62 & 0.43 & 20.82 \\
			$\ell_1$-SVD   & \cellcolor{black!10} 0.088 & 6.35 & 8.36 & 14.33 & 9.36  & 7.60  & 5.63 & 80.8 & 85.4 & 0.62 & 20.04 & 0.58 & 12.38 \\
			SPE-CNN        & 0.000 & 5.53 & 8.08 & 18.79 & 10.54 & 8.84  & 5.55 & 80.6 & 84.9 & \underline{0.39} & 29.47 & \underline{0.32} & 25.99 \\
			Learning-SPICE & 0.000 & \textbf{4.47} & \underline{6.34} & 15.09 & 9.07  & \underline{6.43}  & \textbf{4.50} & \underline{87.4} & \textbf{89.1} & \textbf{0.31} & 18.98 & \textbf{0.28} & 13.82 \\
			ASL-2          & \cellcolor{black!10} 0.049 & 11.62 & 14.28 & 31.66 & 15.57 & 19.91 & 11.06 & 54.6 & 60.7 & 0.77 & 60.00 & 0.62 & 30.00 \\
			SubspaceNet    & 0.000 & 5.15 & 6.79 & \underline{12.71} & \underline{8.88}  & 6.47  & 5.16 & 85.8 & 88.3 & 0.52 & \underline{16.30} & 0.48 & \underline{12.27} \\
			\midrule
			TransDOA (ours) & 0.000 & \underline{4.68} & \textbf{5.99} & \textbf{9.65} & \textbf{7.73} & \textbf{5.18} & \underline{4.69} & \textbf{88.2} & \underline{89.0} & 0.46 & \textbf{11.60} & 0.45 & \textbf{10.93} \\
			\bottomrule
		\end{tabular}%
	}
\end{table*}

Scenario 1 establishes a standard benchmark for multi-source estimation for a ULA. Scenario 2 considers a high-density source environment where the number of signals is near the degrees of freedom of the array.

Scenario 3 represents a scenario with more array elements, which aims to achieve precise DOA estimation. Scenario 4 involves the use of a UCA for two-dimensional localization. This scenario evaluates the model's capability to extend from one-dimensional to two-dimensional tasks.

To ensure a comprehensive evaluation, we performed simulations across a range of angular configurations, SNRs, and snapshots in each scenario. The specific parameter settings are detailed in Table \ref{tab:scenarios_vars}.

\textit{2) Implementation Details:} The proposed TransDOA model comprises 6 Transformer encoder layers. the embedding dimension $D$ is set to 768 and each MHSA module employs 12 attention heads. Within each encoder, the expansion factor of the feed-forward network is set to 4.

Under each simulation scenario, a total of 50,000 training samples and 20,000 validation samples are generated. For the ULA, the DOAs \(\theta\) are randomly generated within the range \([-60^\circ, 60^\circ]\) in the training and validation sets. And for the UCA, the azimuth angle \(\theta\) and elevation angle \(\phi\) are randomly generated within \([-180^\circ, 180^\circ]\) and \([0^\circ, 60^\circ]\), respectively.

Model optimization is performed using the Adam optimizer with a learning rate of $10^{-4}$ and exponential decay rates $(\beta_1,\beta_2)=(0.9, 0.999)$. The number of training epochs is set to 500, with an early stopping mechanism triggered if the validation loss fails to improve for 30 epochs.

To ensure a fair and rigorous comparison, all deep learning models are trained using the identical dataset and optimization protocol, thereby maintaining experimental consistency and the comparability of results. For more detailed implementation and configuration specifics, the complete code is available in our open-source repository\footnote{\url{https://github.com/zzb-nice/DOA_est_Master}}, providing further insight into the structure and parameter settings employed.
\subsection{DOA estimation performance based on TransDOA model} \label{section:5.2}

\subsubsection{Simulation results in Scenario 1} To validate the effectiveness of the proposed TransDOA model, we first evaluate its performance in Scenario 1 through a series of targeted experiments.

\noindent \textbf{Visual Performance Demonstration:} To visually evaluate the robustness of various algorithms under challenging signal environments, Fig. \ref{result_snr_-5_snap_10} illustrates the DOA estimation results in Scenario 1 with SNR= -5 dB and $T$=10, which serves as the low SNR, limited snapshot scenario. The right-hand plot in the figure depicts the absolute angular error of each method. As illustrated, the proposed method demonstrates superior performance, with most estimation errors remaining below \( 25^\circ \), whereas all other algorithms suffer from severe deviations, with absolute errors exceeding 50 degrees.
\begin{figure}[t]
	\begin{center}
		\centering
		\includegraphics[width=\linewidth]{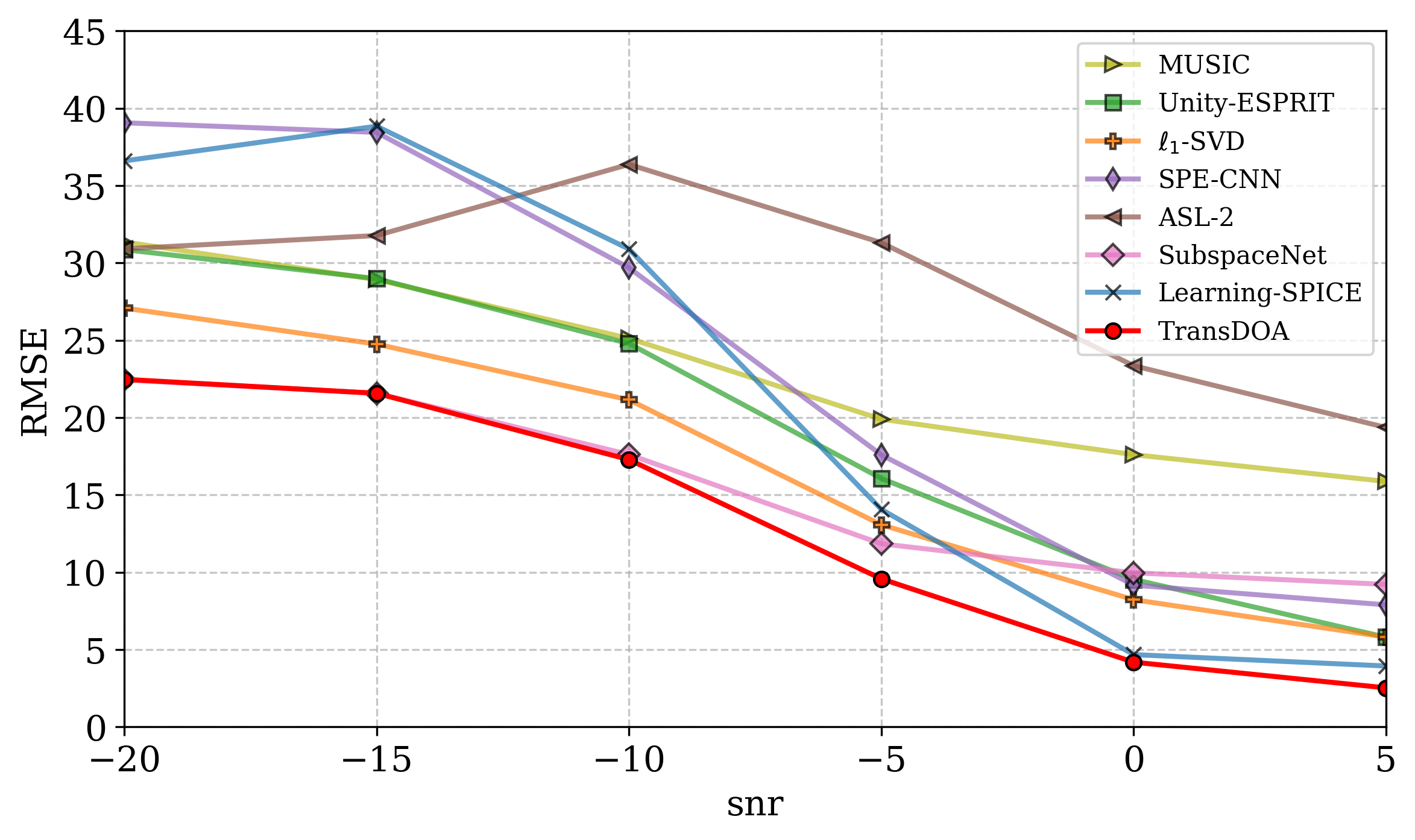}
		\caption{RMSE $vs.$ SNR results under Scenarios 1 with \textbf{Uniform} configuration and \textbf{Params [2]} specified in Table \ref{tab:scenarios_vars}.}
		\label{contrast_loss}	
	\end{center}
\end{figure}

\noindent \textbf{RMSE $\boldsymbol{vs.}$ SNR:} To evaluate the statistical performance of the model, we report the Root Mean Squared Error (RMSE) of the estimated DOAs under varying SNR conditions based on 5,000 simulations. As illustrated in Fig. \ref{contrast_loss}, The proposed method achieves the best RMSE performance when DOAs are randomly generated (\textbf{Uniform} configuration), which can be regarded as the most general case. \textit{Learning-SPICE} closely approaches the performance of the proposed model as SNR increase ($SNR=0db$ or $SNR=5db$). \textit{Unity-ESPRIT} and \textit{SPE-CNN} perform relatively well at high SNRs, while \textit{SubspaceNet} achieves better results in low SNR scenarios. \textit{$\ell_1$-SVD} exhibits consistently stable performance, and \textit{MUSIC} and \textit{ASL-2} exhibit inferior performance compared with the aforementioned algorithms.

\begin{figure}[t]
	\centering
	\begin{minipage}{0.48\linewidth}  
		\centering
		\includegraphics[width=\linewidth]{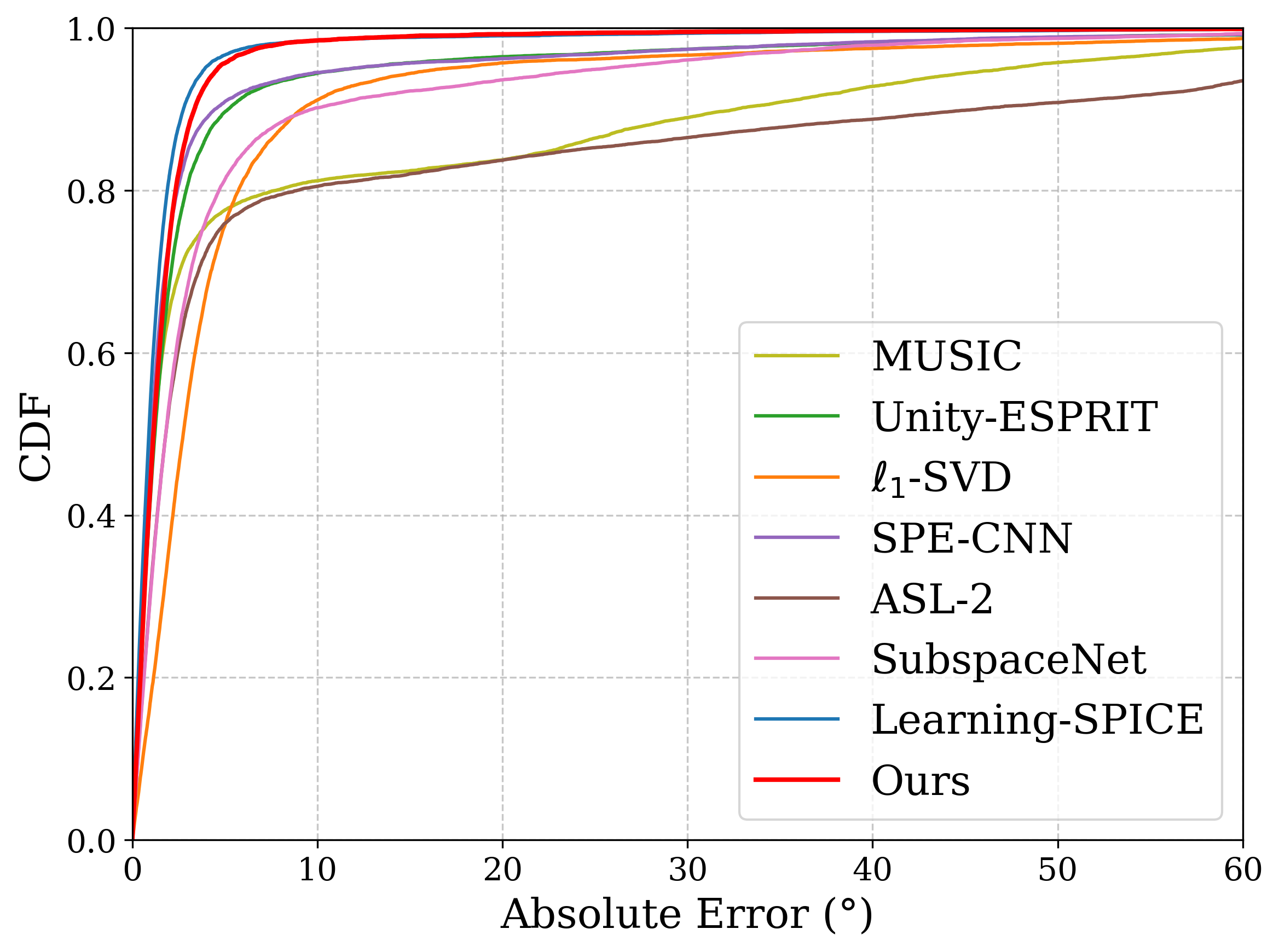}
		\subcaption{$SNR=0db$} \label{cdf_a}
	\end{minipage}
	\hfill
	\begin{minipage}{0.48\linewidth}  
		\centering
		\includegraphics[width=\linewidth]{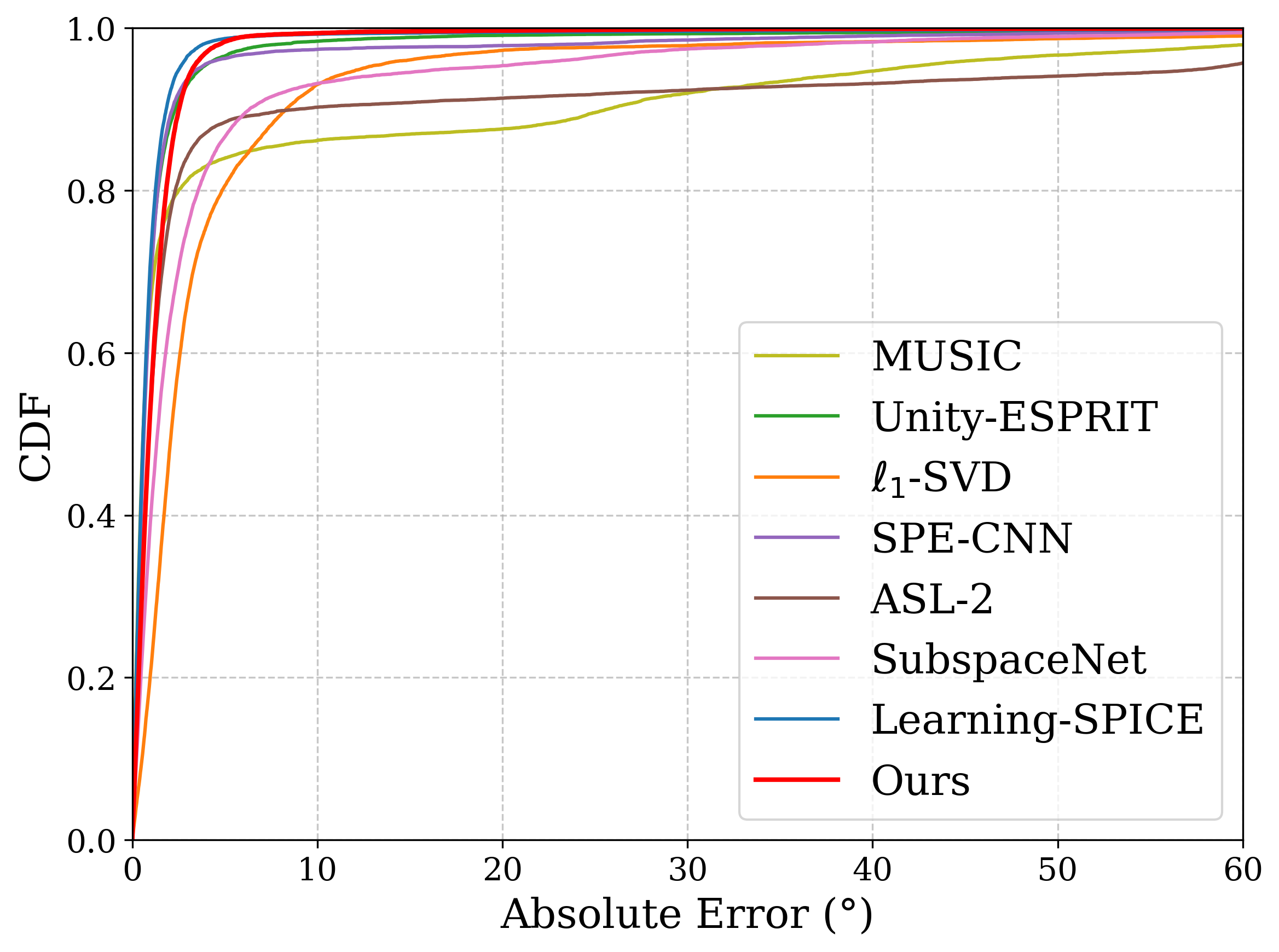}
		\subcaption{$SNR=5db$} \label{cdf_b}
	\end{minipage}
	
	\vspace{0.25cm}  
	
	\begin{minipage}{0.48\linewidth}  
		\centering
		\includegraphics[width=\linewidth]{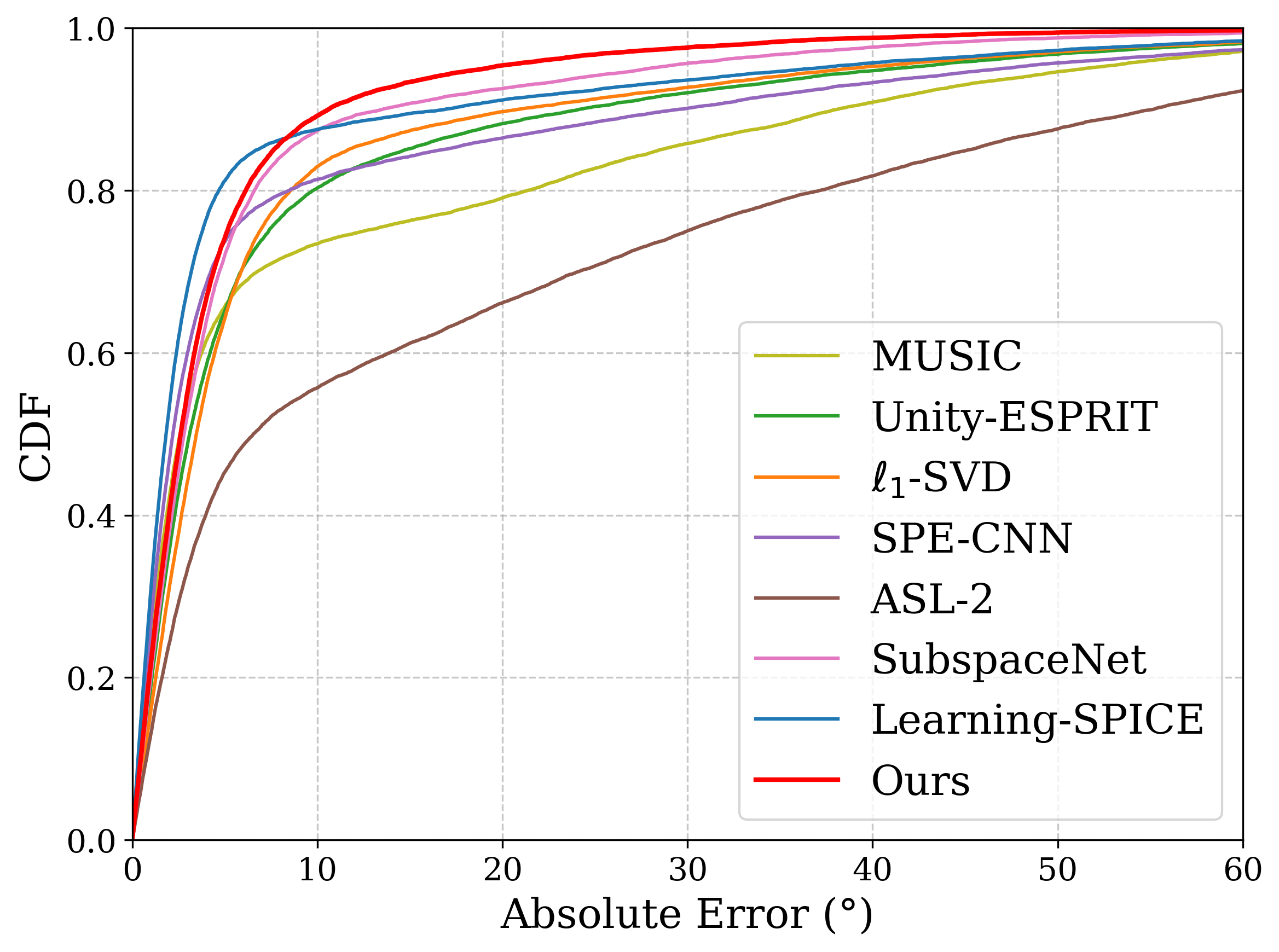}
		\subcaption{$SNR=-5db$} \label{cdf_c}
	\end{minipage}
	\hfill
	\begin{minipage}{0.48\linewidth}  
		\centering
		\includegraphics[width=\linewidth]{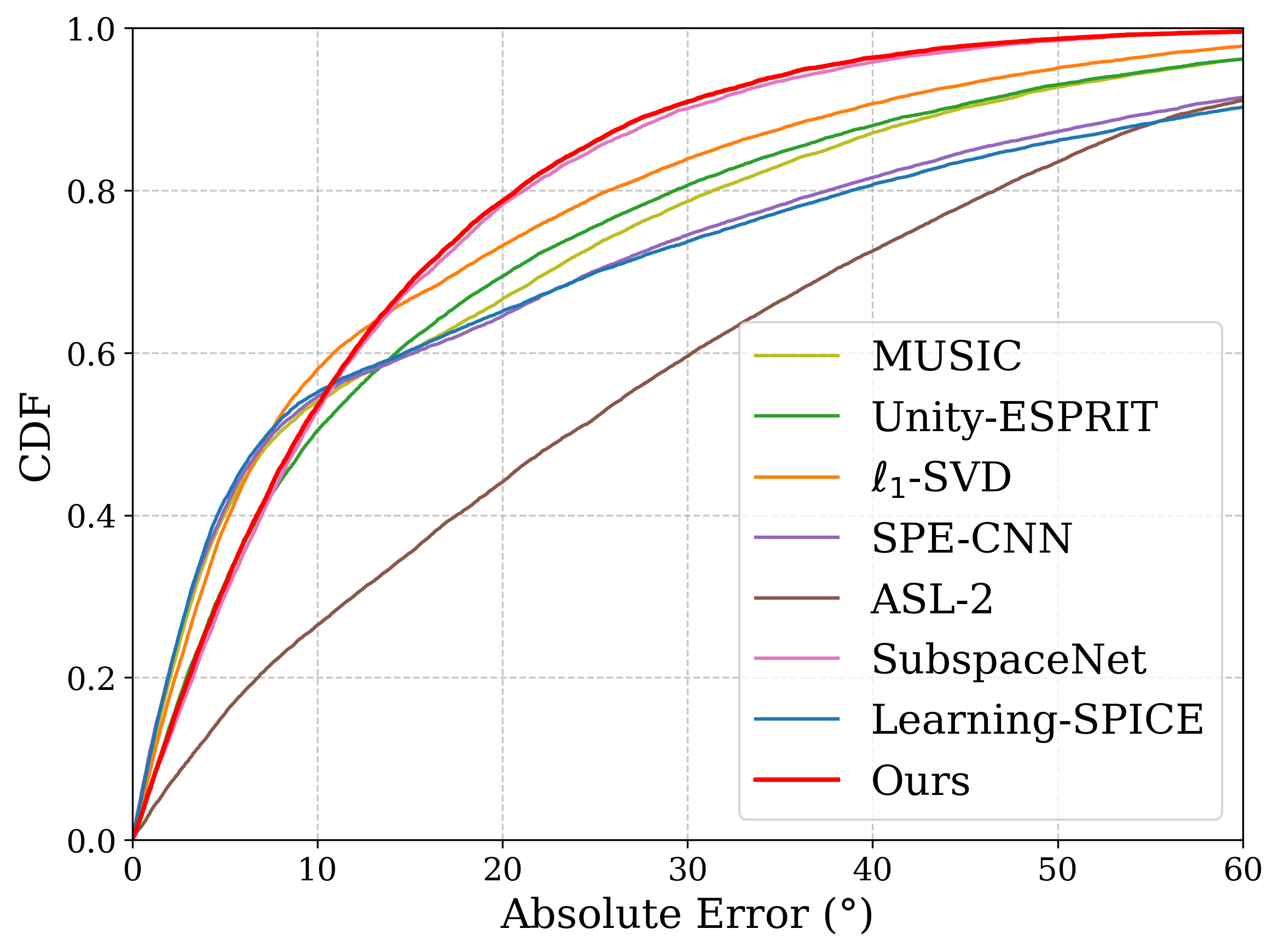}
		\subcaption{$SNR=-10db$} \label{cdf_d}
	\end{minipage}
	
	\caption{ECDF of the absolute error under Scenarios 1 with with \textbf{Uniform} configuration and \textbf{Params [2]}, for (a) $SNR=0db$, (b) $SNR=5db$, (c) $SNR=-5db$ and (d) $SNR=-10db$.}
	\label{cdf_main}
\end{figure}

To provide a more granular evaluation of the statistical characteristics, the empirical cumulative distribution function (ECDF) of the absolute DOA estimation error is illustrated to assess the model's performance. As shown in Fig. \ref{cdf_main}, The ECDF provides insight into the error distribution across 5,000 simulations under Scenarios 1 with \textbf{Uniform} configuration.

As observed from the ECDF curves, while the proposed model exhibits a slightly lower cumulative probability within the high-precision regime, it demonstrates a distinct advantage in larger error ranges. This indicates that the proposed method prioritizes global robustness, making it significantly less prone to catastrophic estimation failures. This trend is particularly evident in low SNR scenarios (e.g., $SNR\leq 0db$).

A representative scenario is depicted in Fig. \ref{cdf_c}. within the high-precision regime ($0^\circ-3^\circ$), the cumulative probability of the proposed model is slightly lower than that of \textit{Learning-SPICE}, \textit{SPE-CNN} and \textit{MUSIC}, while exhibiting advantage over the \textit{SubspaceNet}, \textit{Unity-ESPRIT}, and other algorithms. However, the probability of absolute errors exceeding $10^\circ$ is substantially reduced. This lower frequency of high-error occurrences highlights the model's superior reliability and its ability to suppress outliers in challenging environments.

\begin{figure}[t]
	\centering
	\begin{minipage}{0.98\linewidth}
		\centering
		\includegraphics[width=\linewidth]{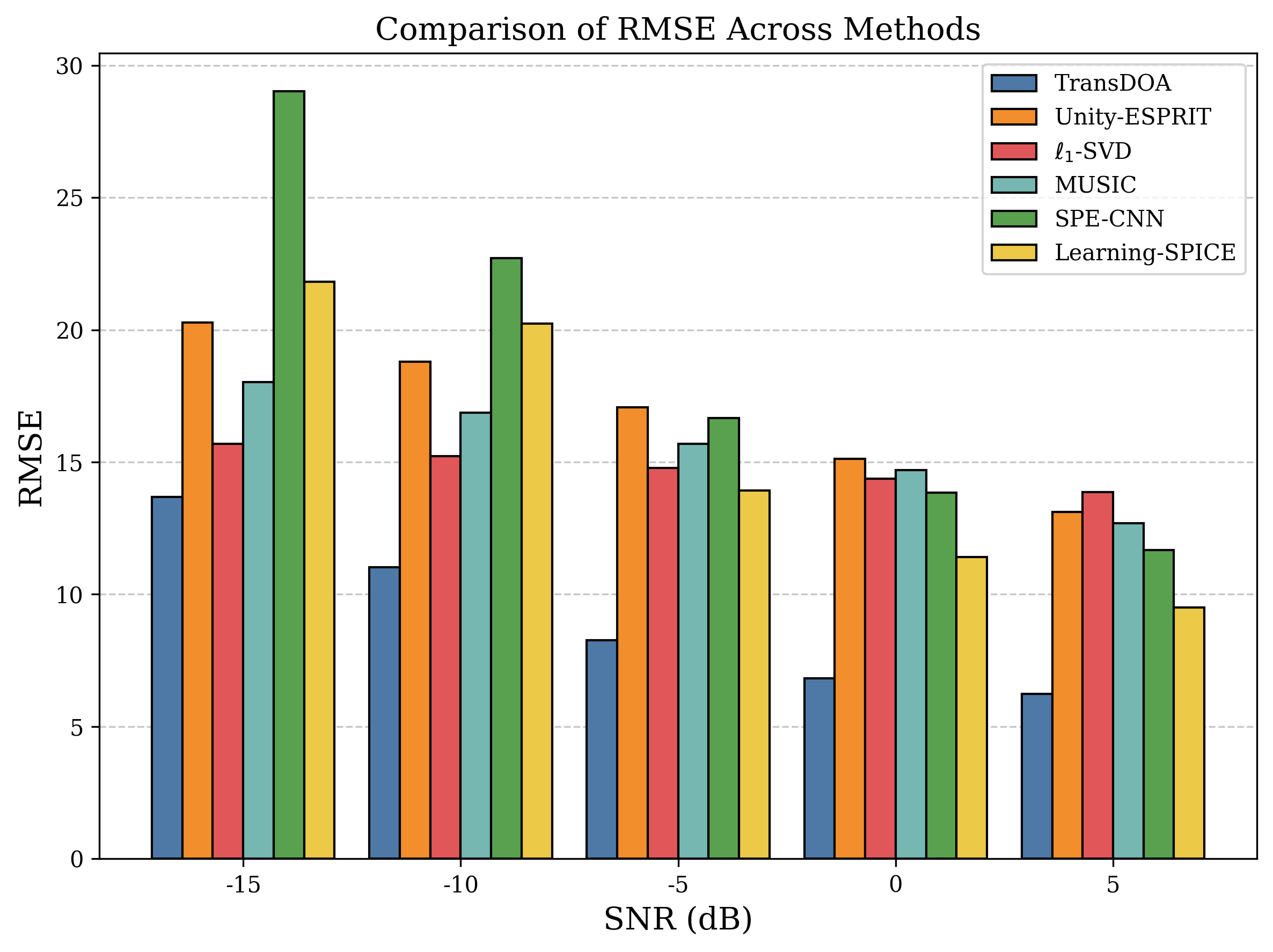}
		\caption{RMSE $vs.$ SNR results under Scenarios 2 with \textbf{Uniform} configuration and \textbf{Params [1]}.}
		\label{bar_result}	
	\end{minipage}
	\hfill
	\begin{minipage}{0.98\linewidth}
		\centering
		\includegraphics[width=\linewidth]{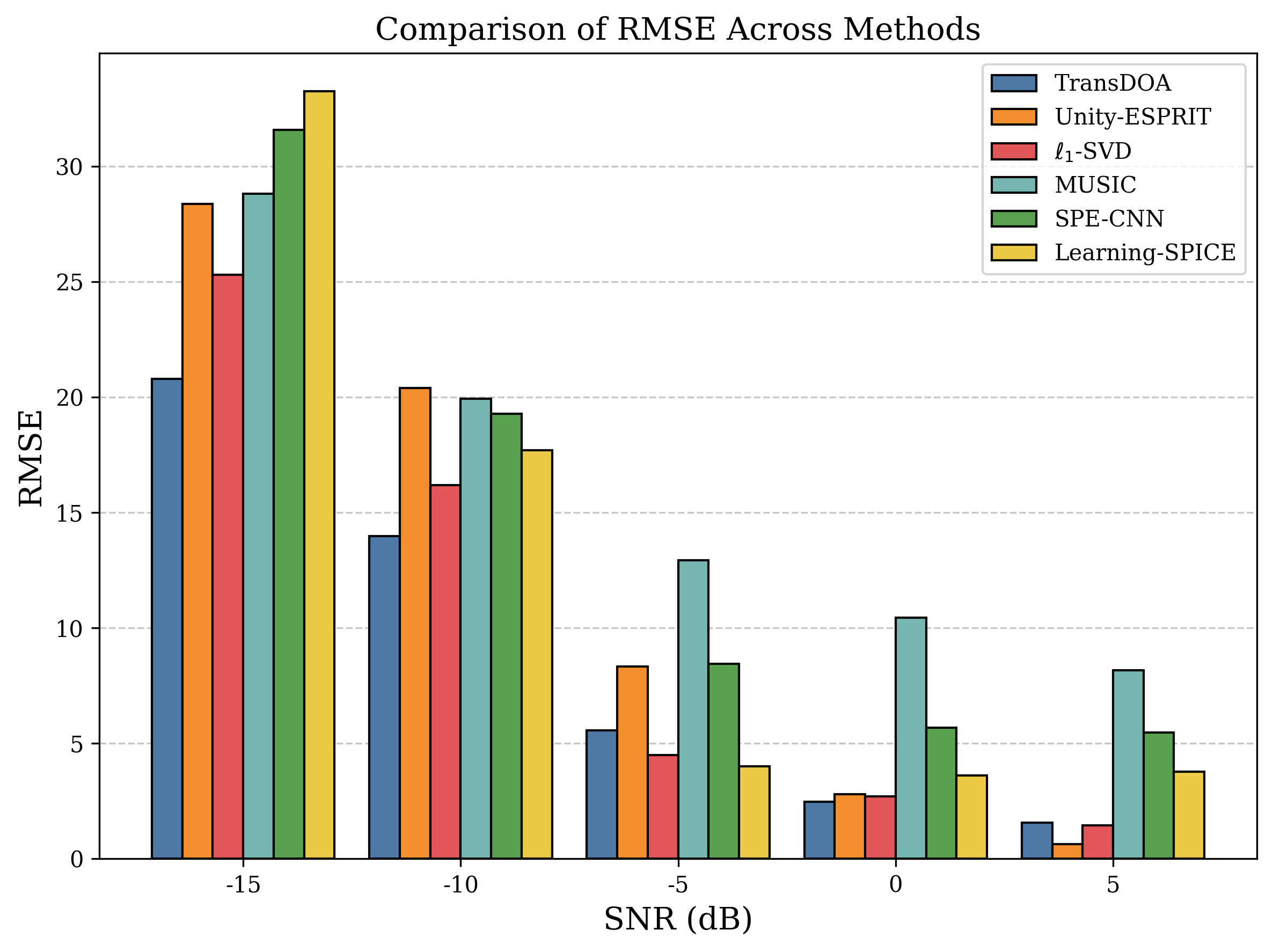}
		\caption{RMSE $vs.$ SNR results under Scenarios 3 with \textbf{Uniform} configuration and \textbf{Params [1]}.}
		\label{bar_result2}	
	\end{minipage}
	
\end{figure}

To rigorously evaluate the model’s performance, we move beyond traditional sorting-based pairing schemes, which fail to accommodate cardinality mismatches and introduce systematic bias by discarding samples with missed detections. Instead, we utilize the Hungarian algorithm for optimal source association, with estimation error capped at $30^\circ$ to prevent extreme outliers from distorting the statistical results. Evaluation is conducted across both 'Raw' and 'Matched' configurations using a comprehensive suite of metrics: OSPA \cite{schuhmacher2008consistent} (to jointly assess cardinality and localization accuracy), MAE, RMSE, Accuracy (Acc., success ratio within a $10^\circ$ tolerance), and the 10th/90th percentiles of the ECDF. These quantiles respectively characterize the high-precision capability and the robustness boundary of the model. 

The performance results for Scenario 1 (Uniform configuration, Params [1]) are detailed in Table \ref{comparison_arrows_moved}. In the table, algorithms with non-zero miss probability are shaded in gray, signifying a potential risk of target loss. Directional arrows ($\downarrow$ / $\uparrow$) are employed to indicate whether a higher or lower value represents superior performance. For clarity, the optimal values for each metric are bolded, while the second-best results are underlined. It is observed that the proposed algorithm consistently achieves the best or second-best performance across the majority of metrics, demonstrating its competitive edge in both estimation accuracy and reliability.

\subsubsection{Simulation results in Scenario 2 and 3}
\begin{figure}[t]
	\centering
	\begin{minipage}{0.48\linewidth}  
		\centering
		\includegraphics[width=\linewidth]{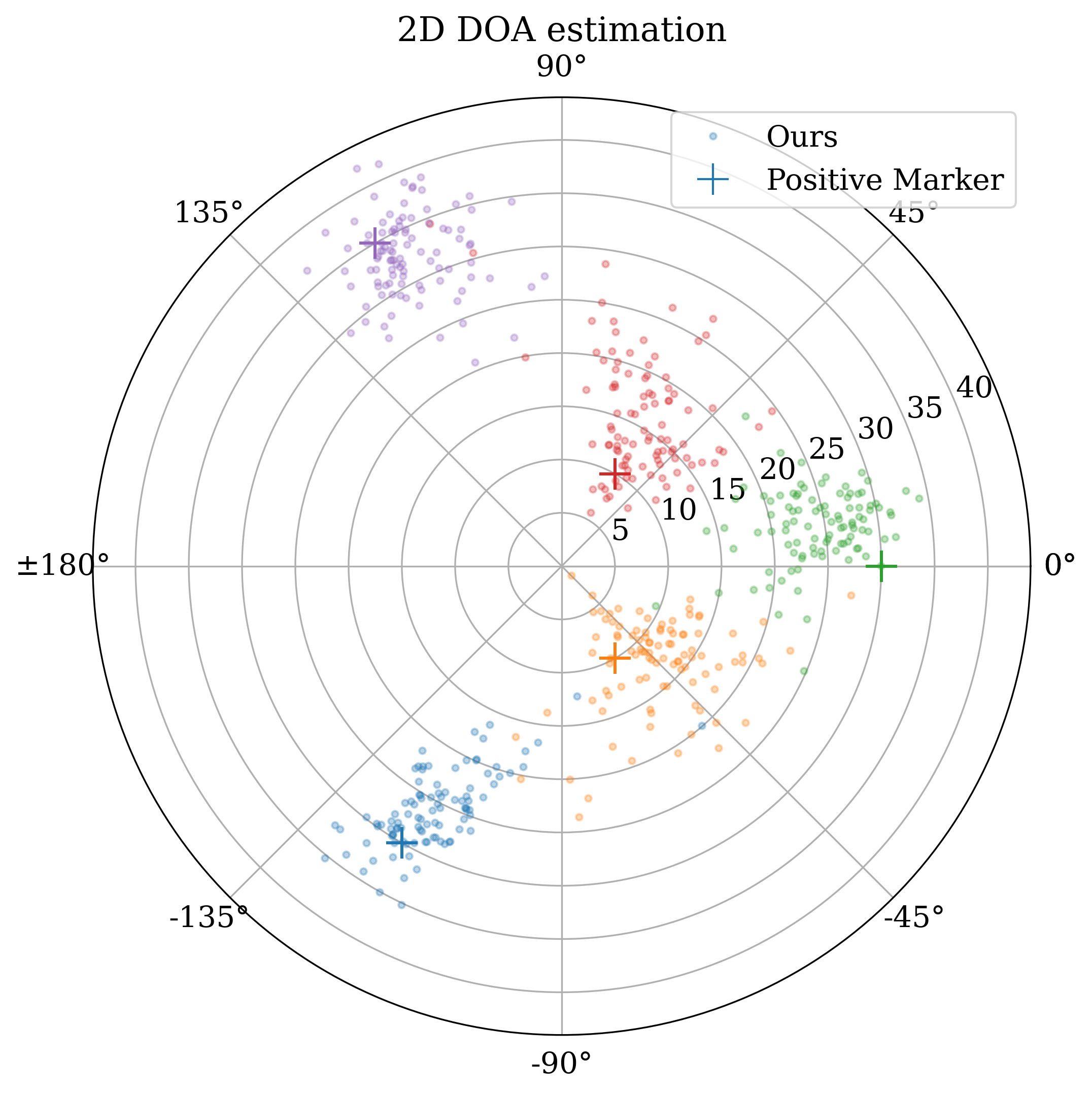}
		\subcaption{TransDOA under Scene 1} \label{uca_a}
	\end{minipage}
	\hfill
	\begin{minipage}{0.48\linewidth}  
		\centering
		\includegraphics[width=\linewidth]{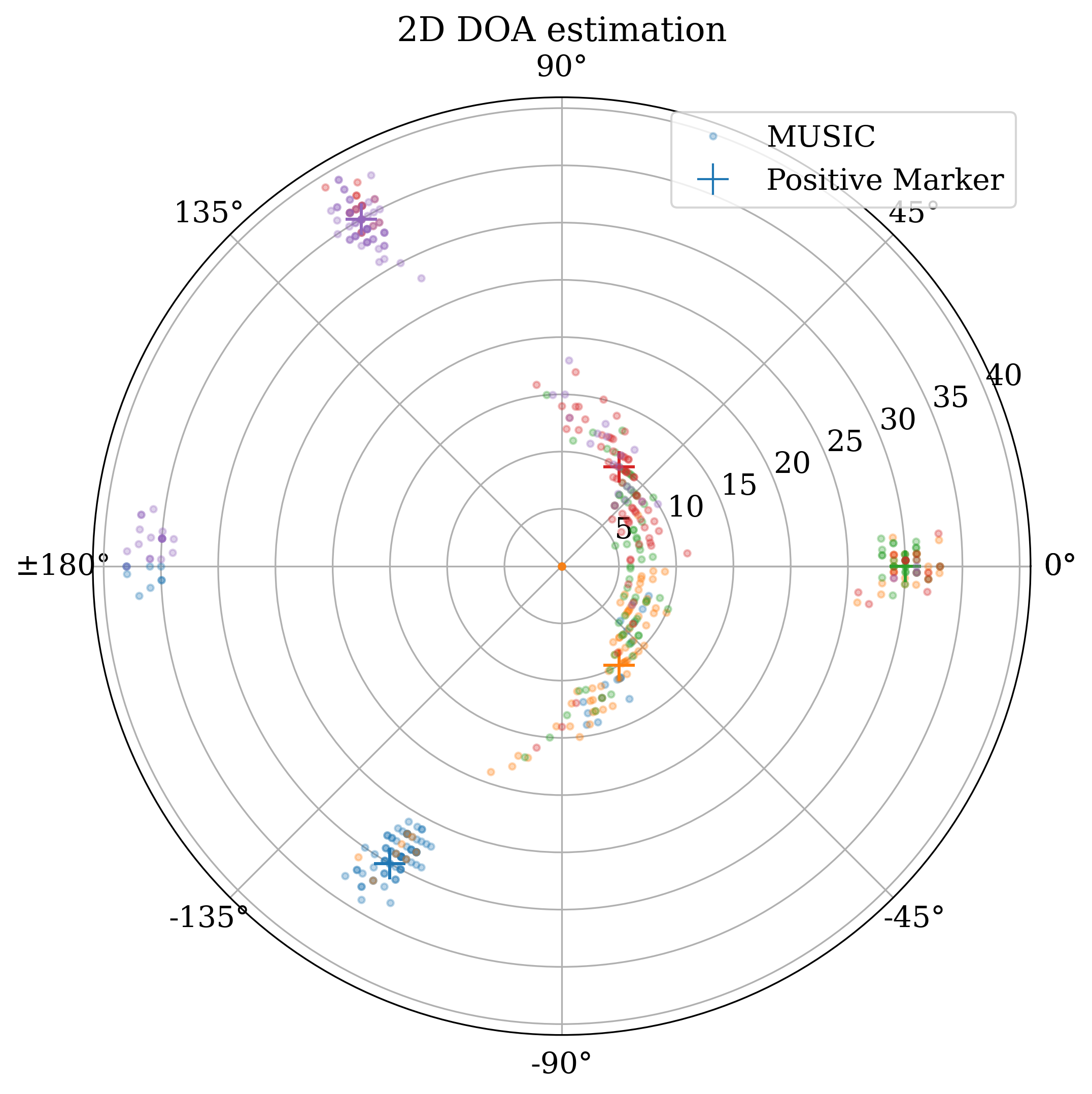}
		\subcaption{\textit{MUSIC} under Scene 1} \label{uca_b}
	\end{minipage}
	
	\vspace{0.25cm}  
	
	\begin{minipage}{0.48\linewidth}  
		\centering
		\includegraphics[width=\linewidth]{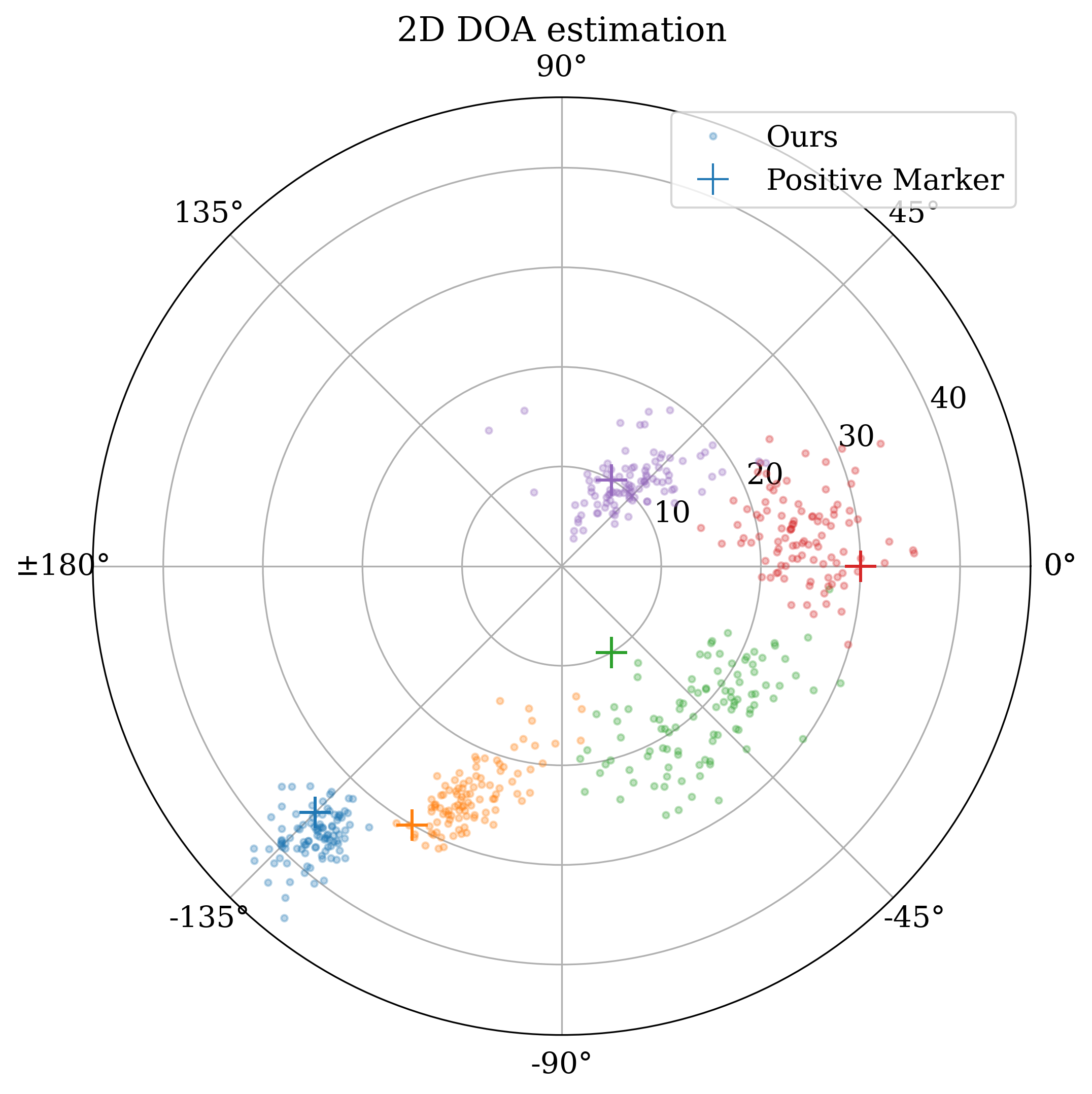}
		\subcaption{TransDOA under Scene 2} \label{uca_c}
	\end{minipage}
	\hfill
	\begin{minipage}{0.48\linewidth}  
		\centering
		\includegraphics[width=\linewidth]{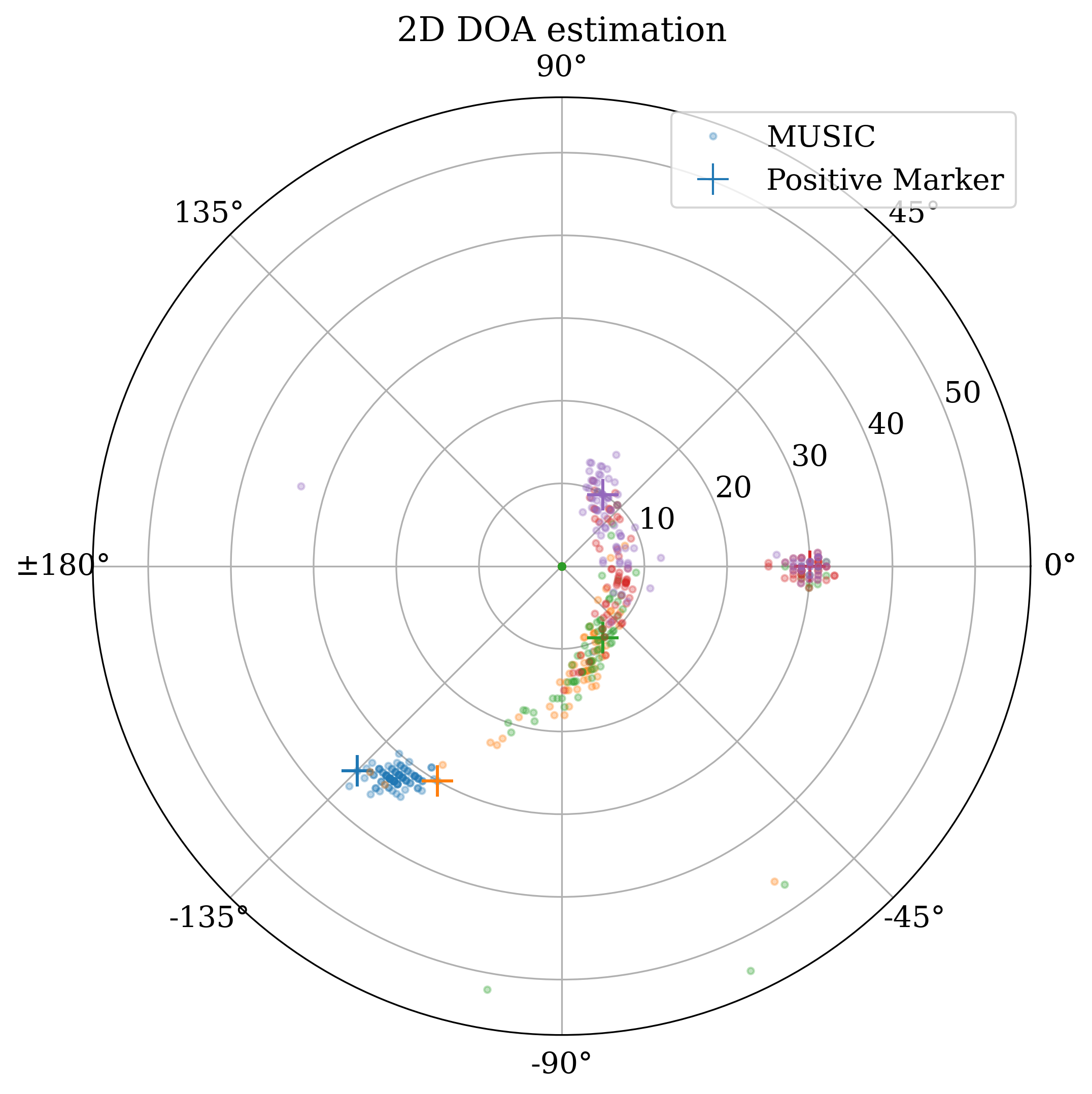}
		\subcaption{\textit{MUSIC} under Scene 2} \label{uca_d}
	\end{minipage}
	
	\caption{Scatter plots of our proposed model and \textit{MUSIC} algorithm under Scenarios 4 with \textbf{Deterministic} configuration and \textbf{Params [1]}.}
	
	\label{uca_radar_main}
\end{figure}

To evaluate the model's performance in handling increased source cardinality and larger array apertures, we conduct further experiments under Scenarios 2 and 3.

The RMSE $vs.$ SNR results under Scenario 2 is illustrated in Fig. \ref{bar_result}. It can be observed that the proposed approach achieves superior statistical performance in low SNR and limited snapshots scenarios compared with all the subspace-based methods, compressive sensing techniques, and state of the art deep learning algorithms. This indicates that the proposed model maintains robustness even in high-density source environment.

Furthermore, the algorithm is evaluated under Scenario 3. As shown in Fig. \ref{bar_result2}, the proposed model outperforms other algorithms in low SNR scenarios, except for cases where $SNR=-5db$ or $SNR=5db$. Due to the limitations of data-driven algorithm fitting accuracy, the proposed model exhibits a performance plateau in the range of 0 dB to 5 dB. Consequently, its performance is inferior to \textit{Unity-ESPRIT} and \textit{$\ell_1$-SVD} algorithms under the $SNR=5db$ scenario, as these analytical methods benefit more effectively from the increased number of array elements. The performance of the proposed model in DOA estimation is difficult to improve in high-precision scenarios, which is an area for future improvement.

\subsubsection{Simulation results in Scenario 4}

To evaluate the model's ability in two-dimensional scenarios,  the proposed architecture is extended to jointly regress the values of elevation $\boldsymbol{\theta}$ and azimuth $\boldsymbol{\phi}$ under Scenario 4.

Similar to one-dimensional DOA estimation, two-dimensional DOA estimation is inherently dependent on multiple factors. Therefore, we consider two DOA configurations with different angular separations. To facilitate a controlled comparison, we only modify one of the angular separation parameters between the two setups. Configuration 1 consists of incident signals at \(\boldsymbol{\theta} = [-120, -60, 0, 60, 120]\) with corresponding azimuth angles given by \(\boldsymbol{\phi} = [30, 10, 30, 10, 35]\). In Configuration 2, the fifth target is relocated to test the resolution limits, resulting in \(\boldsymbol{\theta} = [-135, -120, -60, 0, 60]\) and \(\boldsymbol{\phi} = [35, 30, 10, 30, 10]\).

A representative visualization of the results is presented in Fig. \ref{uca_radar_main}, the symbol "+" represents the true location of the target, while the dot in the figure represents the predicted value of the algorithm. Each predicted value shares the same color as its corresponding ground truth for clarity. Based on 100 Monte Carlo trials per configuration, it can be observed in Configuration 1 that the predicted results of the proposed model are predominantly clustered around the true DOAs. In contrast, while the \textit{MUSIC} algorithm performs adequately for widely separated targets like \(\theta = [-120, 120]\) and \(\phi = [30, 35]\), but it tends to confuse the other three points,  leading to degraded precision. In Configuration 2, the fifth target is moved adjacent to the first. It can be observed that the \textit{MUSIC} algorithm fails to distinguish between the two targets, producing a single merged spectral peak. In contrast, the proposed model maintains high resolution, with predicted values still well clustered around their respective ground truth positions. These results highlight the superior capability of the model in achieving fine-grained localization in complex 2D DOA estimation tasks.
\begin{figure}[t]
	\centering
	\begin{minipage}{0.48\linewidth}  
		\centering
		\includegraphics[width=\linewidth]{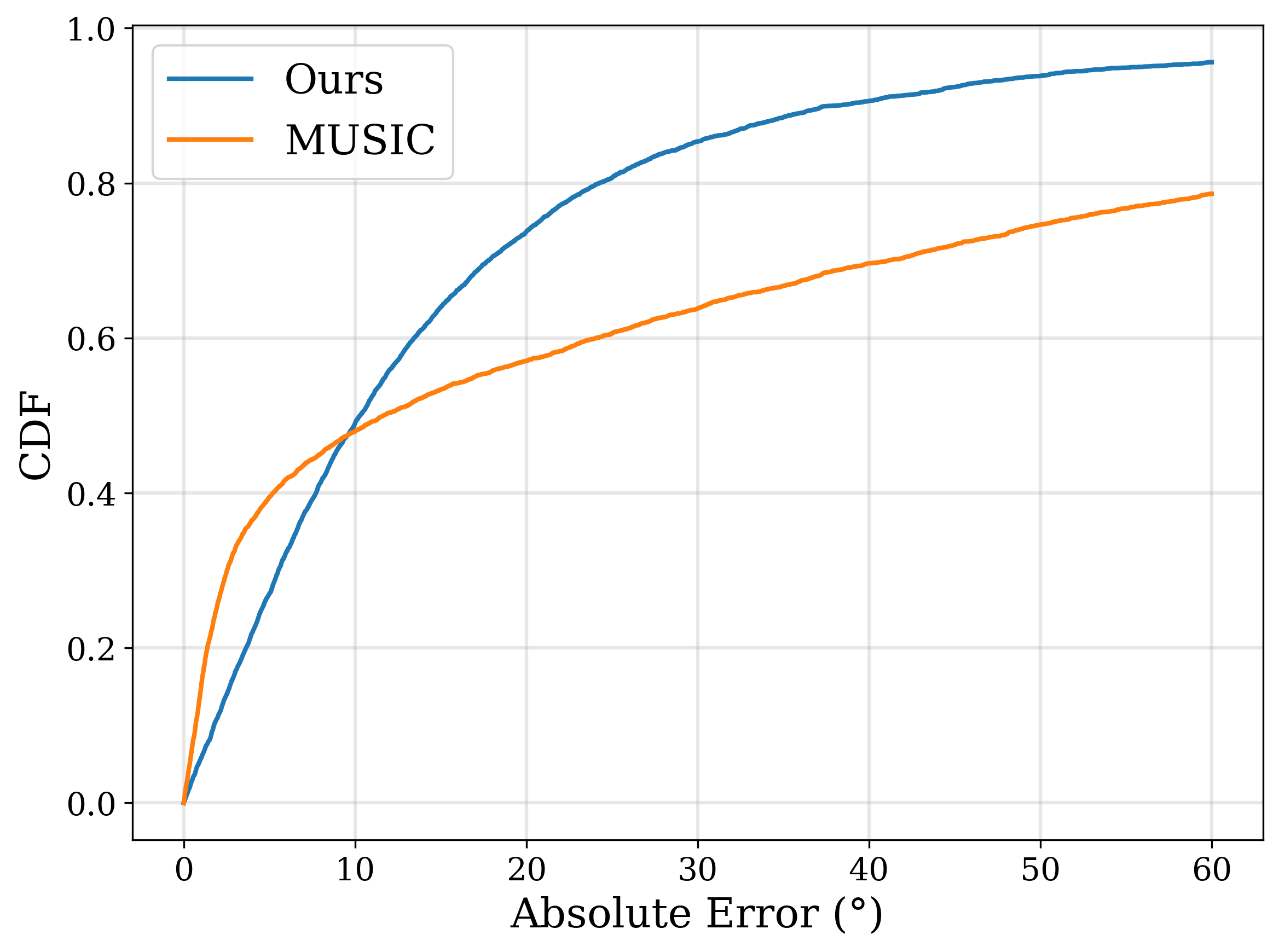}
		\subcaption{CDF of \(\boldsymbol{\theta}\)} \label{cdf_uca_a}
	\end{minipage}
	\hfill
	\begin{minipage}{0.48\linewidth}  
		\centering
		\includegraphics[width=\linewidth]{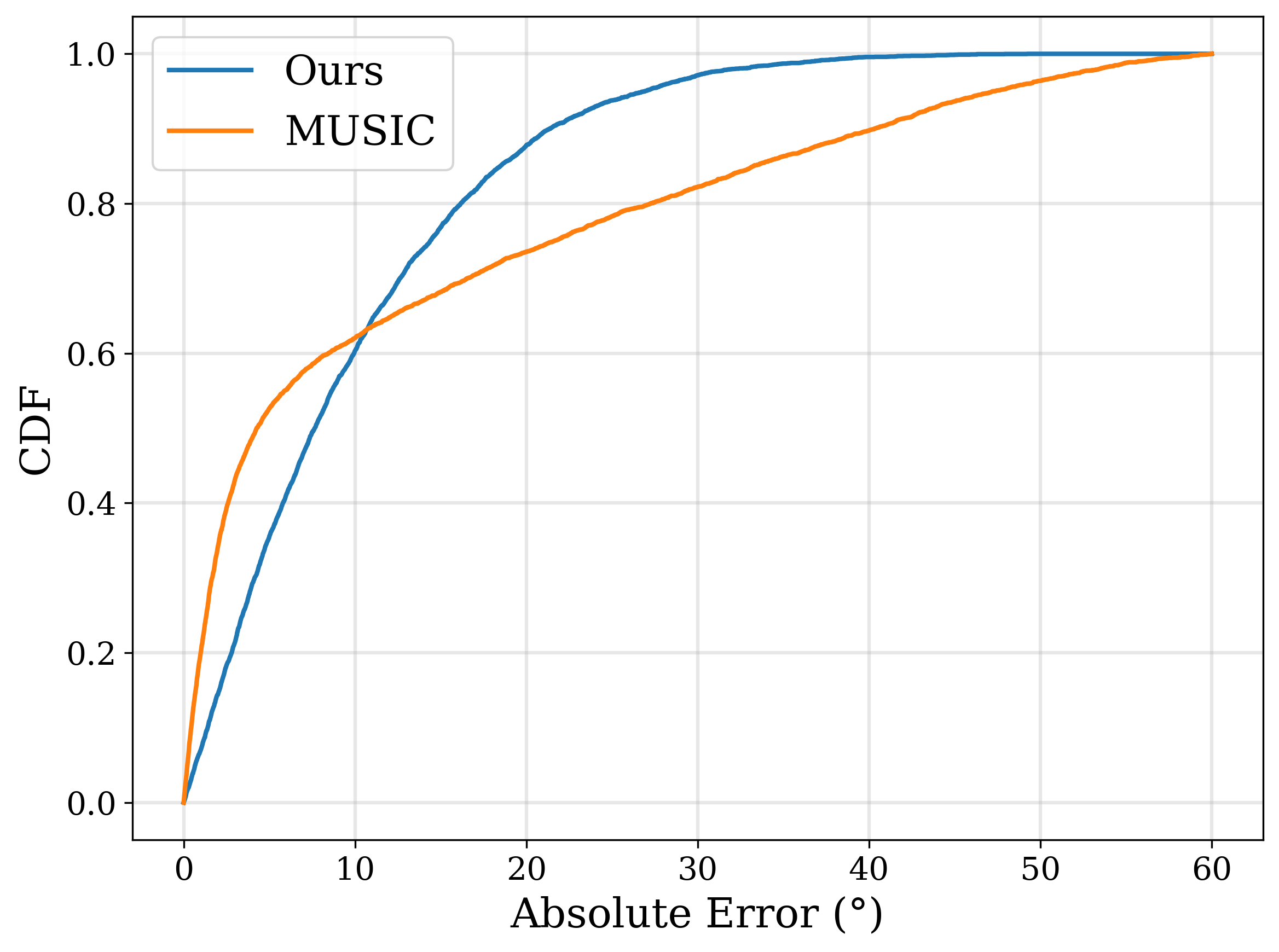}
		\subcaption{CDF of \(\boldsymbol{\phi}\)} \label{cdf_uca_b}
	\end{minipage}
	
	\caption{The CDF of the absolute error under Scenarios 4 with $snap=50$ and $SNR=-5db$ is shown for: (a) \(\boldsymbol{\theta}\), and (b) \(\boldsymbol{\phi}\).}
	\label{cdf_uca_main}
\end{figure}

\begin{figure*}[t]
	\centering
	\begin{minipage}{0.33\textwidth}
		\centering
		\includegraphics[width=\textwidth]{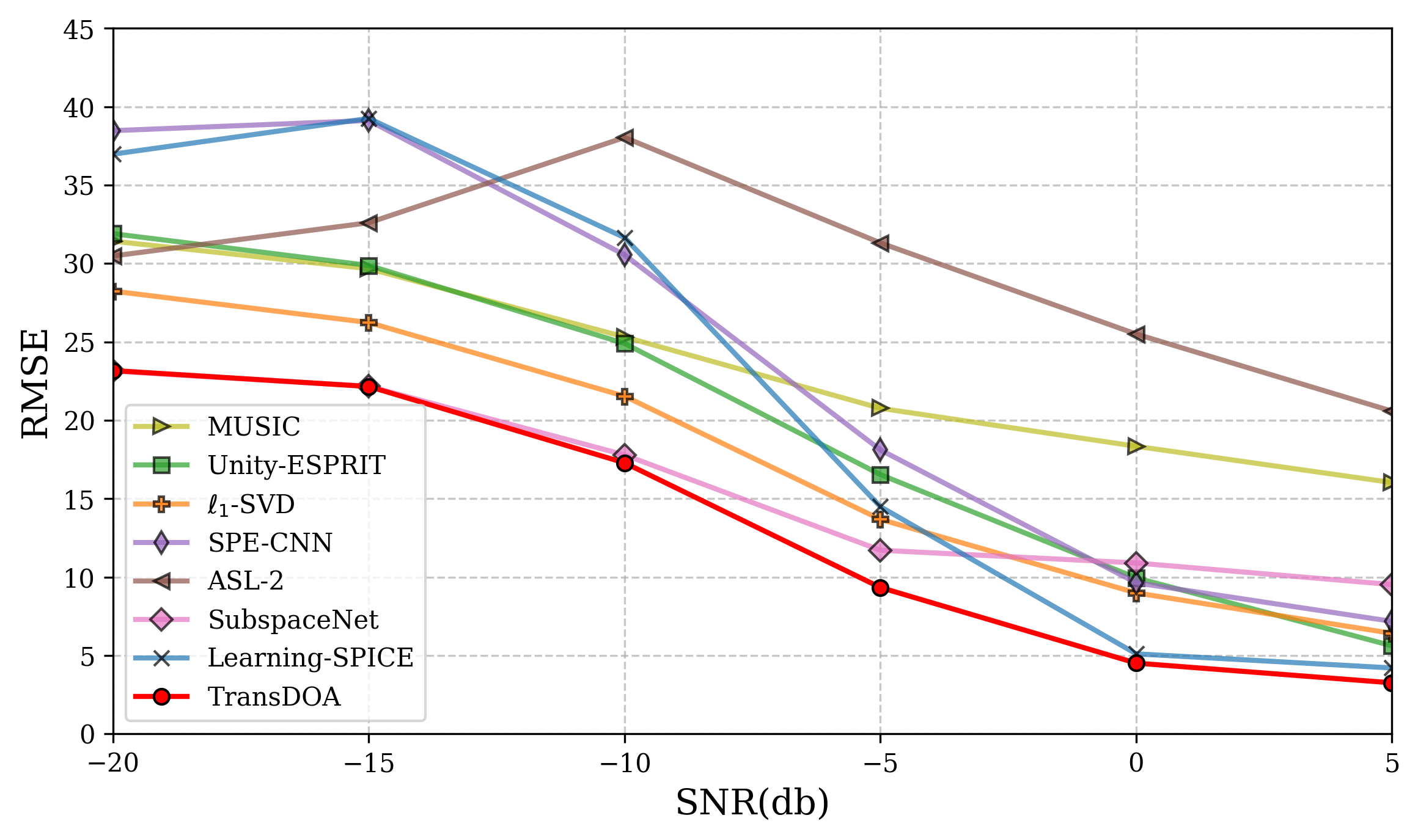}
		\label{array_imperfection_a}
		(a) $\rho=0$
	\end{minipage}%
	\begin{minipage}{0.33\textwidth}
		\centering
		\includegraphics[width=\textwidth]{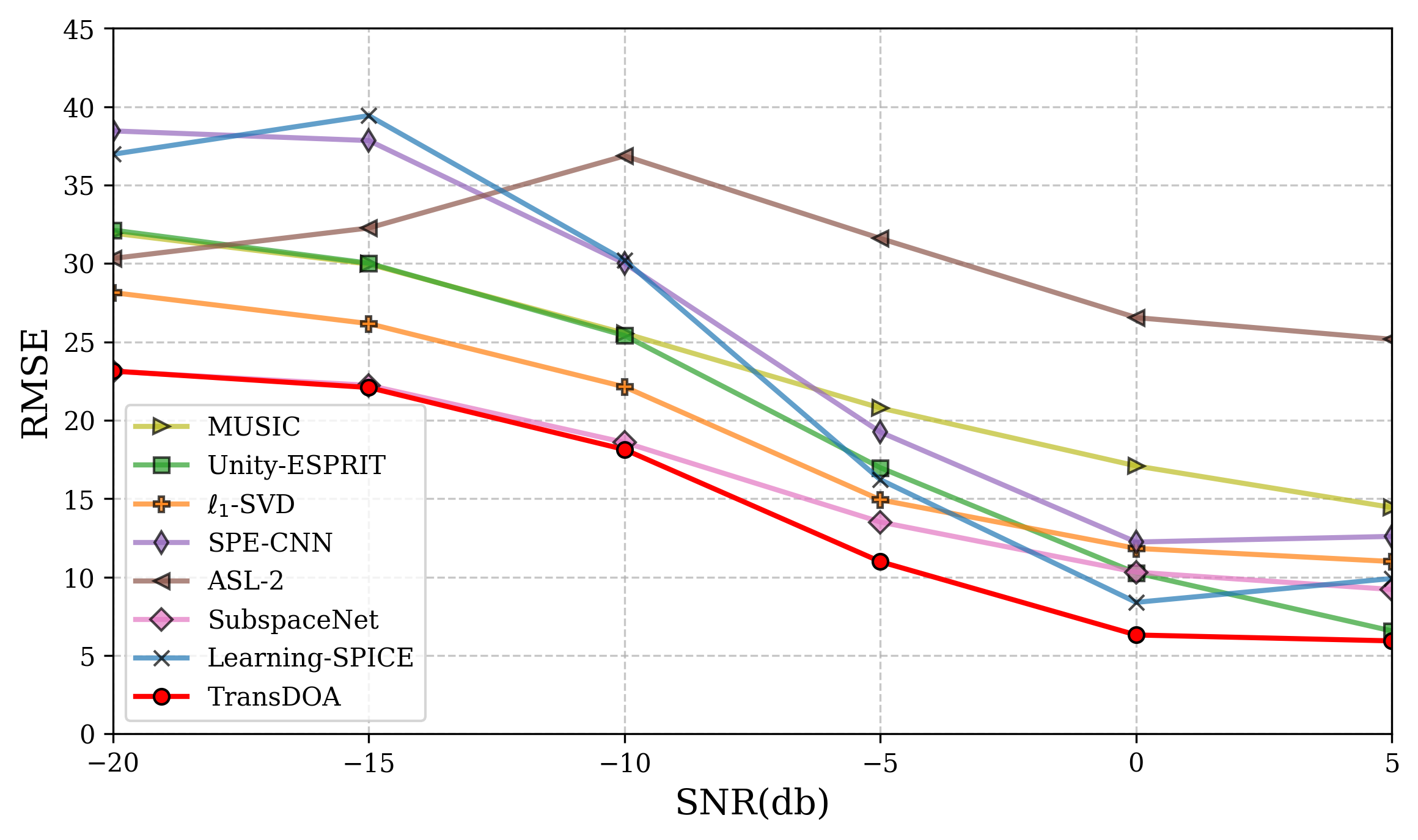}
		\label{array_imperfection_b}
		(b) $\rho=0.5$
	\end{minipage}%
	\begin{minipage}{0.33\textwidth}
		\centering
		\includegraphics[width=\textwidth]{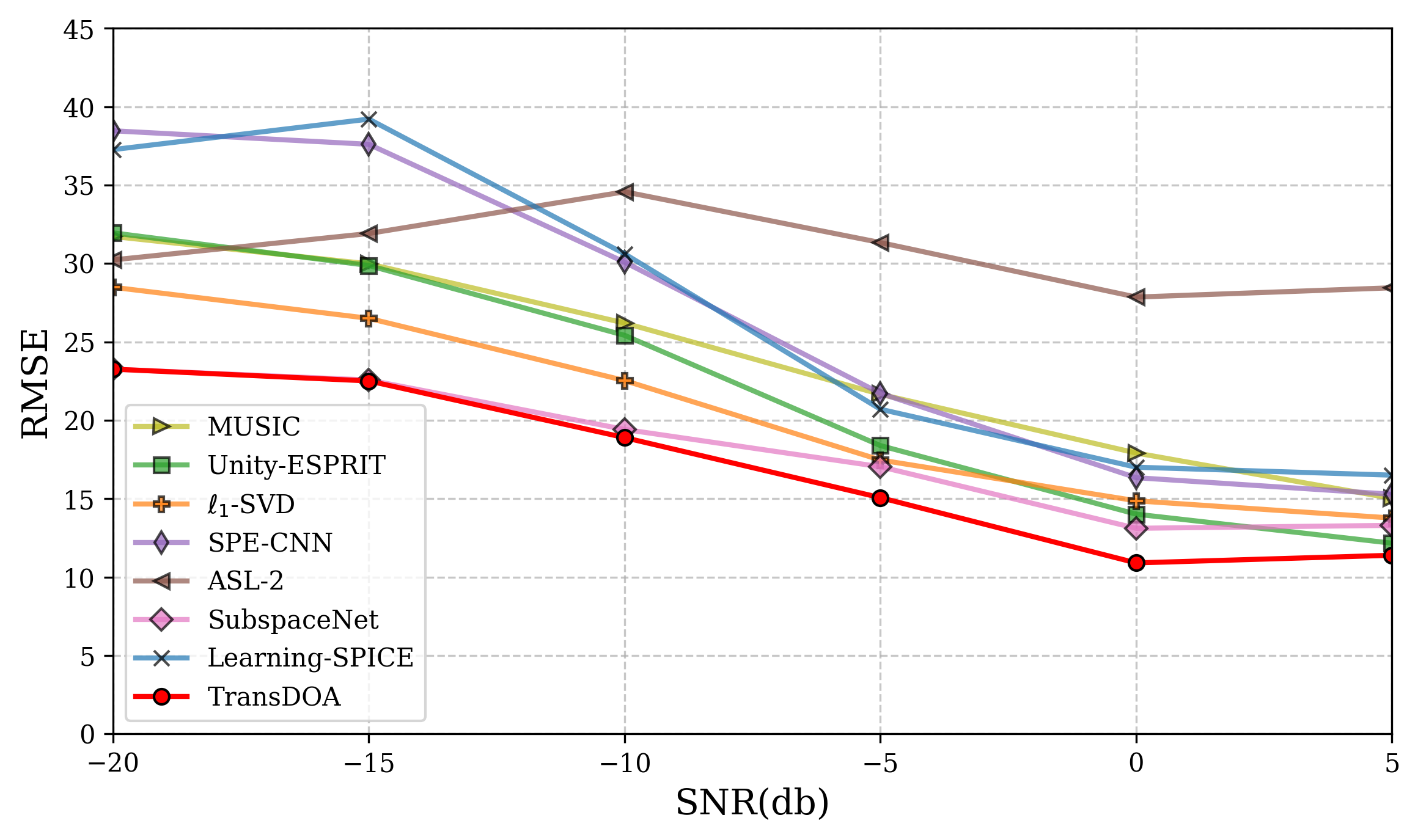}
		\label{array_imperfection_c}
		(c) $\rho=1$
	\end{minipage}
	\caption{RMSE $vs.$ SNR results under Scenarios 1 with \textbf{Uniform} configuration and \textbf{Params [2]}. Three different levels of array imperfections are proposed: (a) $\rho=0$, no array imperfections, (b) $\rho=0.5$, moderate array imperfections, and (c) $\rho=1$, severe array imperfections.}
	\label{array_imperfections}
\end{figure*}

\begin{figure}[t]
	\begin{center}
		\centering
		\includegraphics[width=\linewidth]{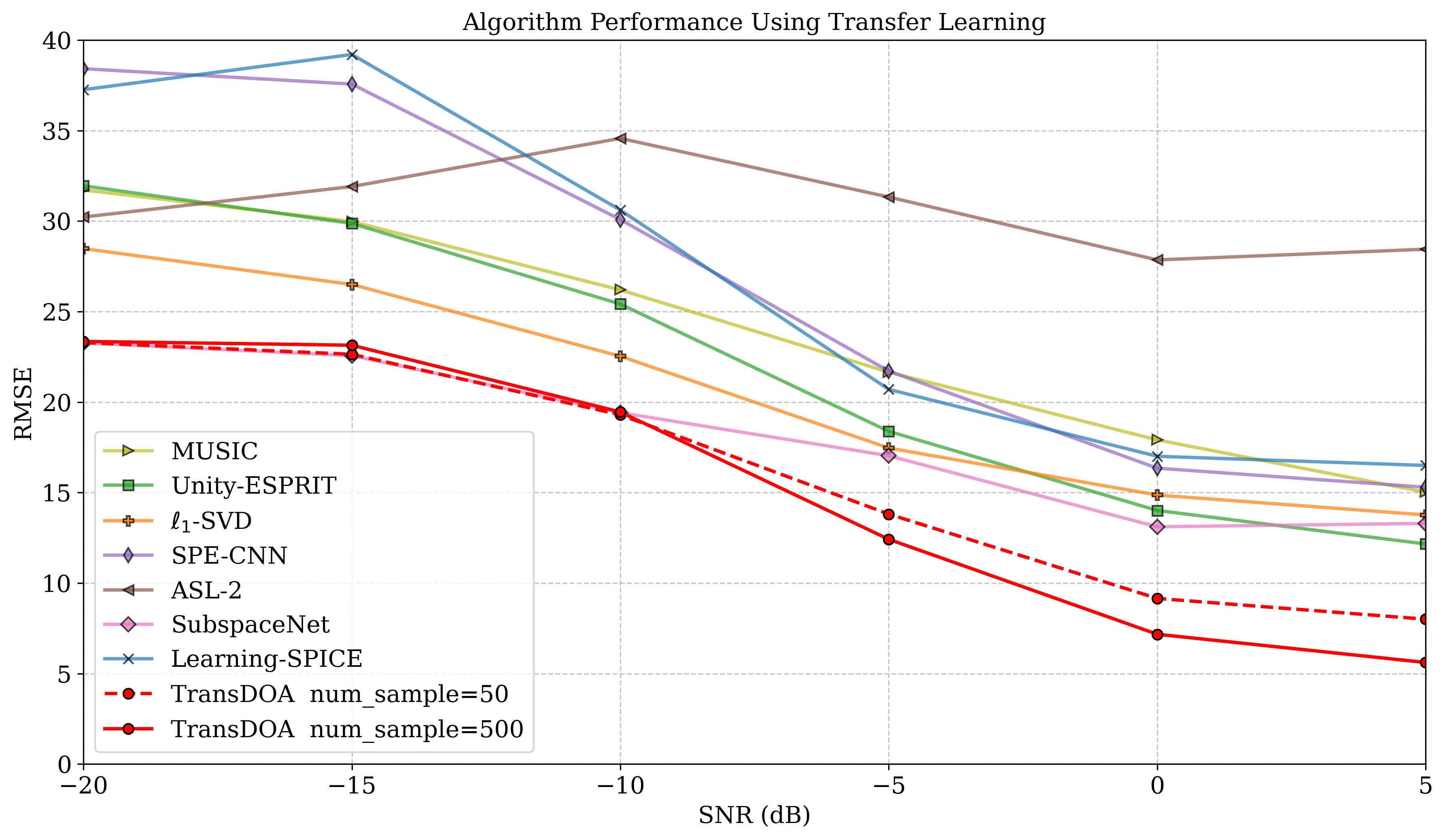}
		\caption{RMSE $vs.$ SNR results under Scenarios 1 with \textbf{Uniform} configuration, \textbf{Params [2]} and $\rho=1$. The dashed line represents the case where only 50 samples are used for transferring learning, while the solid line indicates the use of 500 samples.}
		\label{transfer_learning_performance}	
	\end{center}
\end{figure}
Furthermore, 1000 simulations are performed under Scenario 4 with Uniform configuration and Params [1]. The ECDF of the absolute DOA estimation errors is presented in Fig. \ref{cdf_uca_main}. It can be observed that the MUSIC algorithm performs better on $\text{CDF}(5^\circ)$. However, its performance deteriorates on $\text{CDF}(15^\circ)$. This suggests that although MUSIC can achieve high-precision estimations for certain samples, it is more prone to large estimation outliers. In contrast, the proposed model exhibits superior consistency and robustness.

This experiment further validates the effectiveness of ECDF as a comprehensive metric for DOA estimation. While RMSE remains a critical benchmark, especially when referenced against the CRLB, it only captures the average performance. In comparison, the ECDF provides a more detailed perspective on the error distribution. The results demonstrate that while the proposed model holds a clear advantage in terms of RMSE, the ECDF analysis reveals its subtle limitations in achieving high precision compared to MUSIC's best-case scenarios.

\subsection{Transfer Learning Approach for Array imperfection} \label{section:5.3}
The presence of array imperfections introduces inconsistencies between the received signal model and the ideal model, leading to a degradation in the DOA estimation performance of the model. Scenarios involving multiple array imperfections, particularly under low SNR and limited snapshot conditions, are inherently complex and challenging to handle. In this section, we demonstrate the effectiveness of the proposed transfer learning approach in these scenarios through simulation experiments.

\subsubsection{Transfer Learning in Scenario 1}

\begin{figure*}[t]
	\begin{center}
		\centering
		\includegraphics[width=\linewidth]{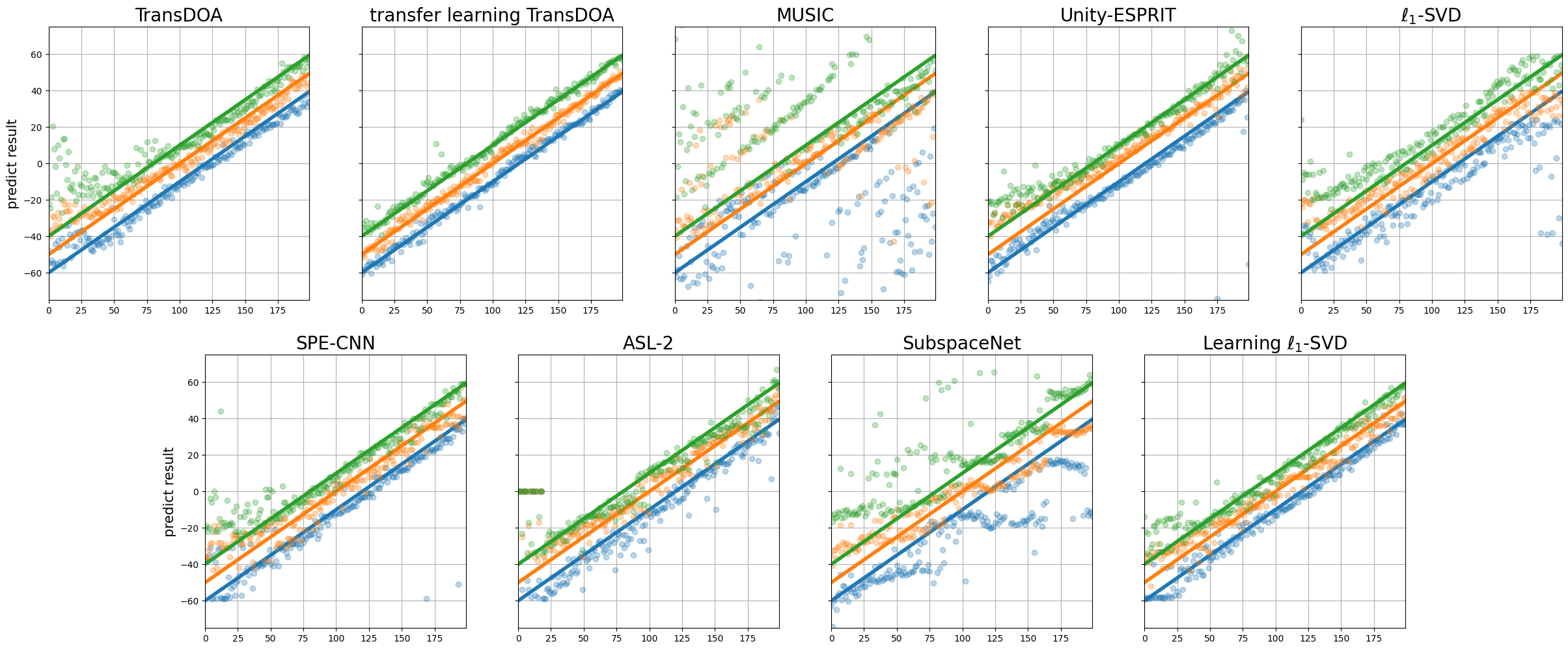}
		\caption{DOA estimation under Scenarios 1 with $SNR=5db$, $snap=10$ and $\rho=1$. The interval is set to $\Delta \hat{\theta} = {10}^{\circ}$. The line represents the true location of the target, while the dot represents the predicted value of the algorithm. }
		\label{pre_result_snr_5_rho_1}	
	\end{center}
\end{figure*}
\begin{table*}[t]
	\caption{Transfer Learning Results with Varying Sample Number}
	\centering
	\begin{tabular}{ccccccccc}
		\toprule
		Sample Number&20&50&100&200&300&500&800&1000 \\
		\midrule
		\midrule
		Directly train&23.03&23.44&23.04&23.03&23.00&14.43&23.00&22.87 \\
		Fine-tune&14.17&14.38&13.57&13.84&13.30&12.79&13.31&12.38 \\
		
		Trnasfer Learning&\textbf{13.66}&\textbf{13.27}&\textbf{13.06}&\textbf{12.72}&\textbf{12.60}&\textbf{12.54}&\textbf{12.21}&\textbf{11.48} \\
		\bottomrule
	\end{tabular}
	\label{transfer_learning}
\end{table*}

\begin{figure}[htbp]
	\centering
	\begin{minipage}{0.48\linewidth}  
		\centering
		\includegraphics[width=\linewidth]{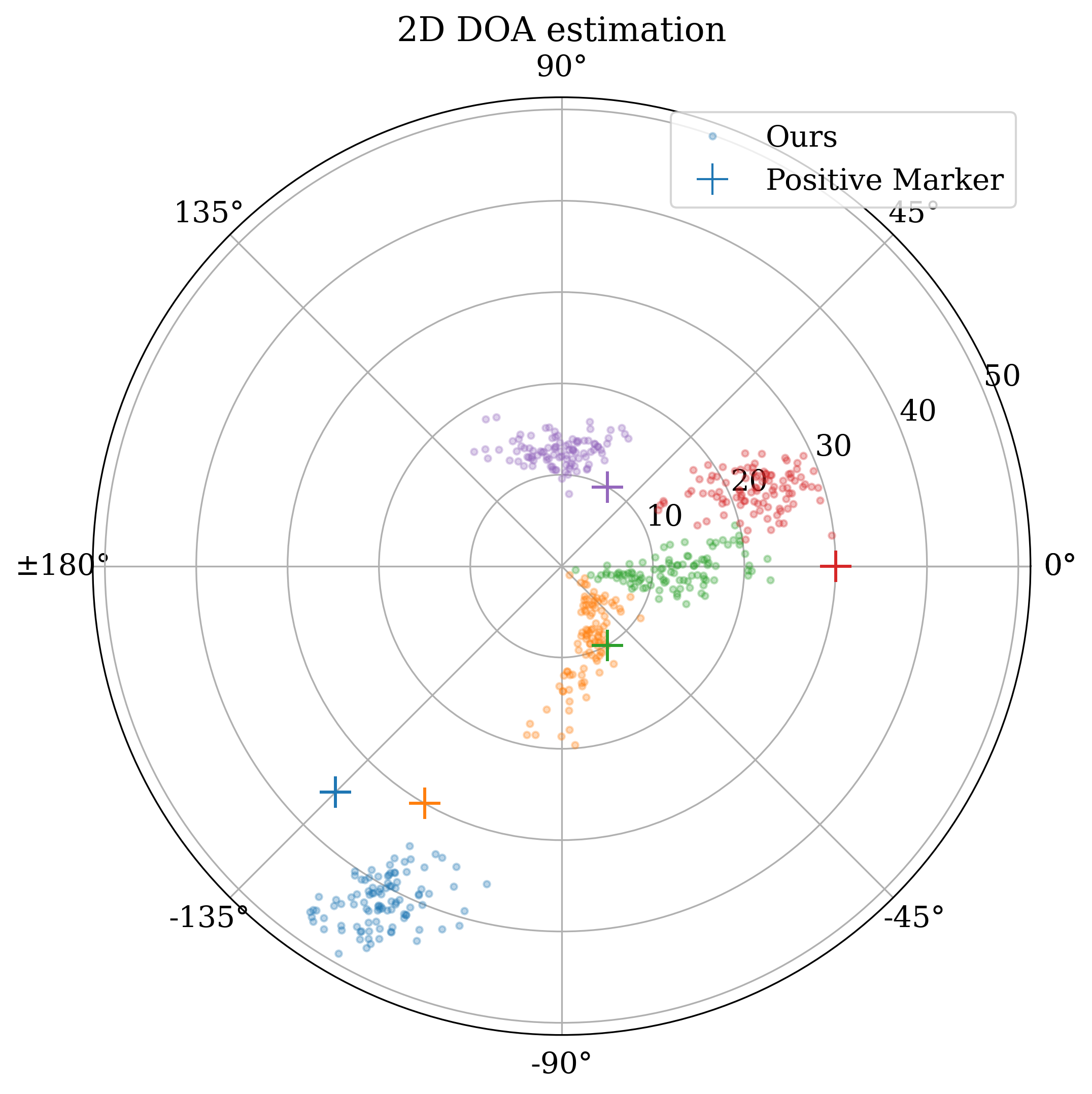}
		\subcaption{TransDOA without transfer learning} \label{2d_transfer_learn1}
	\end{minipage}
	\hfill
	\begin{minipage}{0.48\linewidth}  
		\centering
		\includegraphics[width=\linewidth]{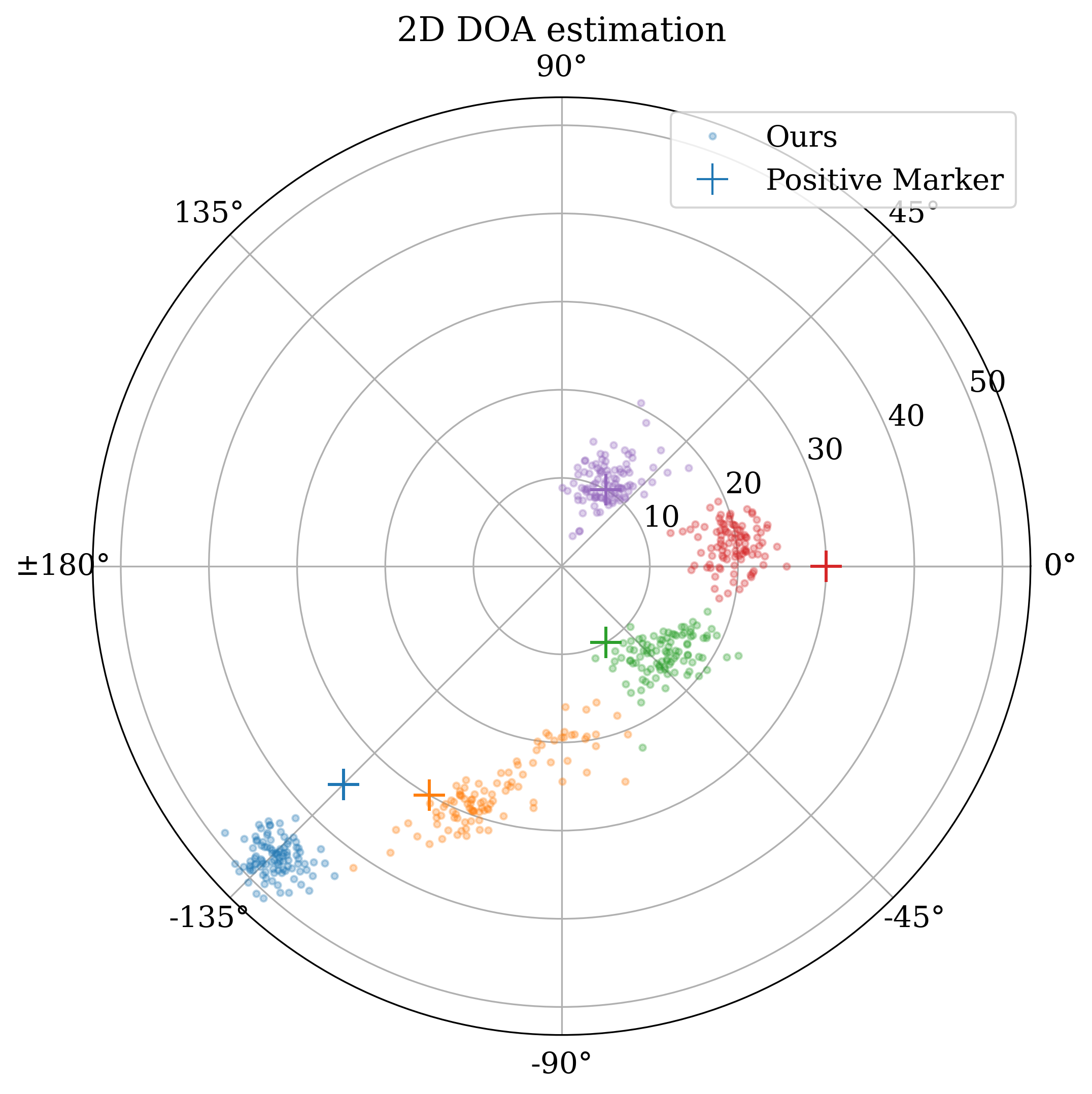}
		\subcaption{TransDOA with transfer learning} \label{2d_transfer_learn2}
	\end{minipage}
	
	\caption{DOA estimation under Scenarios 1 with Equidistant configuration, $SNR=5db$, $snap=10$ and $\rho=1$. 100 Monte-Carlo simulations are conducted and the incident signal is configured with \(\boldsymbol{\theta} = [-135, -120, -60, 0, 60]\), \(\boldsymbol{\phi} = [35, 30, 10, 30, 10]\).}
	\label{2d_transfer_learn}
\end{figure}
To demonstrate the performance degradation of the deep learning model under array imperfections, we illustrate the model's performance under scenarios with multiple array imperfections in Fig. \ref{array_imperfections}. These imperfections include position bias, amplitude and phase inconsistency, and mutual coupling effects, with their overall severity controlled by a hyperparameter \(\rho\). As shown in Fig. \ref{array_imperfections}, It can be observed that under the influence of multiple array imperfections, the statistical performance of deep learning-based methods (the proposed model, \textit{Learning-SPICE} and \textit{SPE-CNN}), compressed sensing-based methods (\textit{$\ell_1$-SVD}), and subspace-based methods (\textit{Unity-ESPRIT} and \textit{MUSIC}) all deteriorates noticeably. Although the proposed model achieves optimal performance under the ideal signal model, particularly in low SNR and limited snapshots scenarios, its performance significantly degrades when subjected to array imperfections. This degradation becomes particularly pronounced in scenarios with relatively higher SNR. When \(\rho = 1\), the performance of the proposed model declines level similar to the traditional algorithms at \(\rho = 0\).

The efficacy of the transfer learning scheme is further validated through 5,000 Monte Carlo trials across varying SNR levels. As shown in Fig. \ref{transfer_learning_performance}, the proposed model achieves significant performance even with only 50 samples, demonstrating accurate DOA estimation under low SNR and limited snapshot scenarios. Moreover, when more samples (500 samples) are used, the model's performance is further enhanced.

To demonstrate the superiority of the proposed transfer learning algorithm, a comparison experiment is carried out under Scenario 1 with $snap=10$, $SNR=-5$ (Params [1]), and array imperfections $\rho=1$. Our method is compared with two other approaches: (1) \textbf{Direct Train}, which involves training from scratch on imperfect data, and (2) \textbf{Fine-tune}, where a pre-trained model on simulated data is adapted samples with array imperfections. The performance of these three methods are evaluated across a varying number of training samples, as summarized in Table \ref{transfer_learning}.  

The results indicate that the proposed transfer learning algorithm consistently achieves superior performance in most cases. This demonstrates that a well-designed transfer learning strategy can utilize the information of limited samples more effectively, outperforming both the Fine-tune and Directly Train methods.  

The impact of transfer learning is further highlighted in Fig. \ref{pre_result_snr_5_rho_1}, considering the Equidistant configuration of Scenario 1 with $SNR=5db$, $snap=10$. In this experiment, three DOAs are generated with a constant interval of $\Delta \hat{\theta} = {10}^{\circ}$, covering the range from -60 to 60 degrees. Evidently, the proposed model exhibits significant bias before applying transfer learning, particularly when the DOAs are close to -60 degrees. Transfer learning effectively rectifies these estimation errors, forcing the predicted estimates to converge tightly around the ground truth. In contrast, the performance of the \textit{MUSIC}, \textit{$\ell_1$-SVD} and \textit{SubspaceNet} suffer from substantial deviations due to array imperfections. Meanwhile, \textit{Unity-ESPRIT} and \textit{ASL-2} fail to maintain reliable estimation as the angular range expands. \textit{Learning-SPICE} and \textit{SPE-CNN}\textit{ASL-2} demonstrate enhanced resilience to array imperfections; however, systematic distortion remains in the predict results.

\subsubsection{Transfer Learning in Scenario 4}

To further validate the scalability of the transfer learning approach, we extend its application from one-dimensional to two-dimensional. Fig. \ref{2d_transfer_learn} presents the DOA estimation results of the TransDOA both before and after applying transfer learning. It can be clearly observed that prior to the application of the transfer learning algorithm, the model fails to effectively distinguish between the first two targets (\(\theta = [-135]\), \(\phi = [35]\) and \(\theta = [-120]\), \(\phi = [30]\)) and suffers from significant bias for other targets due to array imperfections. However, after the transfer learning algorithm is applied, the prediction bias was substantially corrected, and the predicted values were able to converge near the true target.

\section{CONCLUSION}
In this work, we develop a TransDOA model for DOA estimation, which leverages the attention mechanism to capture global dependencies. By directly learning the mapping from the sampled SCM to DOA values, the proposed method can be easily extended to arbitrary array geometries or 2D DOA estimation scenarios, demonstrating strong adaptability. Extensive simulations are conducted to evaluate the performance and robustness of the proposed model. 

Furthermore, deep learning based DOA estimators often suffer from severe performance degradation in practical scenarios with array imperfections. To overcome this limitation, we propose a supervised transfer learning strategy to mitigate the performance degradation in practical scenarios. By aligning the features of the source and target domains in a supervised manner, the proposed method demonstrates improved results in practical scenarios with array imperfections. 

It is worth noticing that this work is the first to introduce an advanced transfer learning strategy to address array imperfections. Investigating how to effectively leverage transfer learning methods to improve the performance of deep learning models in scenarios with array imperfections is a significant and worthwhile research direction, and has the potential to enhance the performance of deep learning algorithms in practical applications. Further exploration of various transfer learning approaches for handling array imperfections is encouraged.

\section*{ACKNOWLEDGMENT}

The authors would like to thank the anonymous reviewers for their insightful comments and suggestions.

\end{document}